\documentclass[10pt,fleqn]{article}

\usepackage{diagbox}
\usepackage{placeins}

    \usepackage{amsmath,amssymb,amsfonts,amsbsy,amsthm}
    \usepackage{latexsym,graphics,graphicx,color,sectsty,mathdots,bm}
    
    \usepackage{subcaption}
    \usepackage{layout}
    \usepackage{bbm}
    \usepackage[colorlinks=true,linkcolor=blue]{hyperref}
    \usepackage{calc,cancel}
    \usepackage{enumitem}
    \usepackage{rotating}
    \usepackage[table]{xcolor}
    \usepackage{cleveref}
    \usepackage{appendix} 
    \usepackage{sidecap}
    \usepackage{amscd}
    \usepackage{mathtools}
    \usepackage{scalerel}

    \usepackage{arydshln}
    \usepackage{makeidx} 
    \usepackage{nicematrix}
    \usepackage{fancyhdr}
    \usepackage{xfrac}
    \usepackage{tocvsec2}

\theoremstyle{plain}
\newtheorem{theorem}{Theorem}

\newtheorem{lemma}[theorem]{Lemma}

\newtheorem{definition}[theorem]{Definition}

\newtheorem*{summary*}{Summary}

\theoremstyle{definition}
\newtheorem{example}[theorem]{Example}

\theoremstyle{remark}

	\definecolor{bgblue}{rgb}{0.04,0.39,0.53}
	\definecolor{dblue}{rgb}{0,0.3,0.7}
	\definecolor{ddblue}{rgb}{0,0.1,0.6}
	\definecolor{ddgreen}{rgb}{0,0.25,0.05}
	\definecolor{dgreen}{rgb}{0,0.5,0.05}

\newcommand{\req}[1]{(\ref{#1.eq})}

\newcommand{\beq}{\begin{equation}}
\newcommand{\eeq}{\end{equation}}
\newcommand{\beqn}{\begin{eqnarray}}
\newcommand{\eeqn}{\end{eqnarray}}
\newcommand{\beqns}{\begin{eqnarray*}}
\newcommand{\eeqns}{\end{eqnarray*}}
\newcommand{\bct}{\begin{center}}
\newcommand{\ect}{\end{center}}
\newcommand{\btmz}{\begin{itemize}}
\newcommand{\etmz}{\end{itemize}}
\newcommand{\benum}{\begin{enumerate}}
\newcommand{\eenum}{\end{enumerate}}

\newcommand{\R}{{\mathbb R}}

\newcommand{\tb}{{\bar{t}}}
\newcommand{\xb}{{\bar{x}}}
\newcommand{\lba}{{\bar{l}}}

\newcommand{\bbm}{\begin{bmatrix}} 
\newcommand{\ebm}{\end{bmatrix}} 

\newcommand{\bsm}{\left[ \begin{smallmatrix}} 
\newcommand{\esm}{\end{smallmatrix} \right]} 

\newcommand{\bsbm}{\left[ \begin{smallbmatrix}} 
\newcommand{\esbm}{\end{smallbmatrix} \right]} 

\newcommand{\bbNm}{\begin{bNiceMatrix}} 				
\newcommand{\ebNm}{\end{bNiceMatrix}} 
\newcommand{\bNA}[1]{ \left[ \begin{NiceArray}{#1} } 		
\newcommand{\eNA}{ \end{NiceArray} \right] }

\newcommand{\lb}{\left(}
\newcommand{\rb}{\right)}
\newcommand{\lcb}{\left\{}
\newcommand{\rcb}{\right\}}

\newcommand{\lbar}{\left|}
\newcommand{\rbar}{\right|}

\newcommand{\cI}{{\sf I}}

\newcommand{\be}{\begin{equation}}
\newcommand{\ee}{\end{equation}}

\newcommand{\cplxs}{ C\kern -.35em \rule{0.03 em}{.7 ex}~   }

\def\complex{\hbox{C\kern -.45em \rule{0.03 em}{1.5 ex}}~}

\newcommand{\rmi}{{\rm i}}

\newcommand{\rmv}{{\rm v}}

\newcommand{\bi}{\begin{itemize}}
\newcommand{\ei}{\end{itemize}}
\newcommand{\ben}{\begin{enumerate}}
\newcommand{\een}{\end{enumerate}}

\newcommand{\cA}{\mathcal{A}}

\newcommand{\bseq}{\begin{subequations}}
\newcommand{\eseq}{\end{subequations}}

\newcommand{\ba}{\begin{array}}
\newcommand{\ea}{\end{array}}

\newcommand{\LP}[1]{{\sf L}^{\!#1}}

\newcommand{\mycaption}[1]{\caption{\footnotesize #1}}
\newcommand{\mysubcaption}[1]{\caption{\footnotesize #1}}

\definecolor{dred}{rgb}{.8,0,0}

\newcommand{\sm}{\text{-}}

\def\clap#1{\hbox to 0pt{\hss#1\hss}}

\newcommand{\btc}{\begin{tabular}{c}}
\newcommand{\btbl}{\begin{tabular}{l}}
\newcommand{\et}{\end{tabular}}

\newcommand{\dm}[2]{{{\sf d}\!\lb #1 , #2 \rb}}

    \newcommand{\wba}{{\bar{w}}}

    \newcommand{\xd}{\dot{x}}

	\newcommand{\rom}{\rule{0em}{1em}}
	
	\newcommand{\romn}{\rule{0em}{.91em}}
	\newcommand{\rome}{\rule{0em}{.85em}}

\newcommand{\hsom}{\hspace{1em}} 
\newcommand{\hstm}{\hspace{2em}}

\newcommand{\step}[1]{\mathfrak{h}\!\left( #1 \right)}
\newcommand{\heavi}{\mathfrak{h}}

\newcommand{\rms}{{\rm s}}

\newcommand{\sfX}{{\sf X}}

\newcommand{\txtcm}[1]{\mbox{\tiny\textcircled{#1}}}

\newcommand{\Volt}{{\cal V}}

\newcommand{\smint}[2]{\scaleobj{.8}{\int_{{#1}}^{{#2}}}}

\newcommand{\blb}{\big(}
\newcommand{\brb}{\big)} 

\newcommand{\llb}{{\underline{l}}}
\newcommand{\lub}{{\bar{l}}}

\newcommand{\lbb}{\big(}
\newcommand{\rbb}{\big)}

	\newcommand{\bbms}{\begin{bsmallmatrix}}
	\newcommand{\ebms}{\end{bsmallmatrix}}

               \DeclareMathAlphabet{\mymathbb}{U}{BOONDOX-ds}{m}{n}

\theoremstyle{definition}
\setlist[itemize]{leftmargin=*}

\newcommand{\FIGLOC}{figures}

\theoremstyle{definition}
\newtheorem{ExInternal}{Exercise}

\title{A Tutorial on Solution Properties of State Space Models of Dynamical Systems}

\author{Bassam Bamieh\thanks{Department of Mechanical Engineering, University of California at Santa Barbara, {\em bamieh@ucsb.edu.} This work is partially supported by NSF Awards CMMI-1763064 and ECCS-1932777.}}

\date{}

\begin{document}

\maketitle

\begin{abstract} 
		The starting point of analysis of state space models is investigating existence, uniqueness 
		and solution properties such as the semigroup property, and various formulas for the solutions. 
		Several concepts such as the state transition matrix, the matrix 
		exponential, the variations of constants formula (the Cauchy formula), 
		 the Peano-Baker series, and the Picard iteration are used to characterize solutions. 
		 In this note, a tutorial treatment is given where all of these concepts are shown to be 
		 various manifestations of a single abstract method, namely solving equations using 
		 an operator Neumann series involving the Volterra operator of forward integration. 
		 The matrix exponential, the Peano-Baker series, the Picard iteration, and the Cauchy formula 
		  can be ``discovered'' naturally from this Neumann series. 
		 The convergence of the  series and iterations is a consequence of the key property of 
		  asymptotic nilpotence of the Volterra operator. This property  is an asymptotic version 
		 of the nilpotence property of a strictly-lower-triangular matrix. 
\end{abstract}

%


\section{Introduction}

	State space models are the starting point in analysis of dynamical systems. They come in various forms 
	of generality as follows 	
	\be
            	\begin{aligned} 
			(a):~ 
            		\xd(t) ~&=~ A\big( x(t),t\big) , 
		\hstm\hstm 
            			&	
			(b):~
            		\xd(t) ~&=~ A\big( x(t) , u(t) \big) , \\ 
            		(c): ~ 
			\xd(t) ~&=~ A(t) ~ x(t), 
            			& 	
			(d):~
            		\xd(t) ~&=~ A(t) ~x(t) ~+~ B(t) ~u(t) , 		\\
            		(e): ~ 
			\xd(t) ~&=~ A ~ x(t), 
            			& 	
			(f):~
            		\xd(t) ~&=~ A ~x(t) ~+~ B ~u(t) . 		
            	  \label{SS_non.eq}
            	\end{aligned} 
	\ee
	The state  at each time $x(t)\in\R^n$ is an $n$-vector, while the input $u(t)\in\R^q$ is also a vector at each 
	$t$, with typically a different dimension than the state. For control problems, for example,  the signal $u$ 
	is the control input, and most interesting problems have the dimension of $u$ being much less than that of $x$
	(controlling many states with a single or few inputs). If the signal $u$ is a disturbance or  a noise signal, it 
	typically has dimensions comparable to those of the state $x$. 
	
	The first column in~\req{SS_non} represents systems without an external input, and we generally want to understand their 
	responses $\lcb x(t) \rcb$  due to various boundary conditions $x(\tb)$ specified at some time $\tb$. 
	In the second column, the signal $u$ is regarded as an external 
	signal, and we typically want to establish response properties for a {\em whole class} of inputs $u$ rather than 
	a single, fixed input. 
	
	The systems in the first row are generally nonlinear, and without making more restrictive 
	assumptions on the structure 
	of the vector field $A$, one can only deduce rather basic properties of existence and uniqueness of solutions. 
	An important special instance of (\textit{a})  is the time-invariant case where $A$ is constant in $t$. 
	The second row consists of linear time-varying systems. We will be able to say  more about them, 
	but in general these are capable of very rich behavior, and again without additional restrictive assumptions, only 
	basic properties can be established. The third row represents linear time-invariant systems, and much more can 
	be said about properties of those systems. Those statements will generally involve linear-algebraic properties 
	of the matrices $A$ and $B$. 
	As a side note, when the state dimension becomes very large or infinite, 
	 the distinctions between the three categories of systems listed above can become quite 
	blurry and in some cases cease to be relevant. 
	
		This tutorial is motivated by and organized around  a pedagogical principle that it is better to discover results 
		starting from basic, generally applicable principles than to simply be told what the answer is,  and then just verify it. 
		I will try to illustrate what I mean by this using the most concrete case of linear time-invariant systems. 
		The traditional treatment~\cite{kailath1980linear,antsaklis1997linear,hespanha2018linear} 
		to derive formulas for the solution of a linear time-invariant state space 
		system
	\be
		\xd(t) ~=~ Ax(t) ~+~ w(t), 
		\hstm \hstm x(0)=\xb
	 \label{ss_PB.eq} 
	\ee
	 proceeds as follows. 
	 First, the homogenous problem with $w(t)=0$ is addressed. The solution of this problem 
	 is given in terms of the {\em matrix exponential}. 	 
	 Given any square matrix $A$,  the exponential function is defined  by the series 
	formula 
	\[
		e^{At} ~:=~ \sum_{k=0}^\infty \frac{A^k t^k}{k!} ~=~ I + At + \frac{1}{2} A^2  t^2+ \frac{1}{3!} A^3 t^3~+~ \cdots 
	\]
	It is not difficult  to show that this series is absolutely convergent for any matrix $A$ and time $t$. 
	Note also that $e^{0}=I$. 
	By differentiating this series element by 
	element, it then follows that the derivative of this matrix-valued function is the matrix-valued function 
	 \[
	 	\frac{d}{dt} e^{At}  ~=~ A ~e^{At} ~=~ e^{At} ~A. 
	 \]
	 It is then an easy verification that the solution of~\req{ss_PB} (with $w=0$) is given by  
	 \be
	 \begin{aligned} 
	 	x(t) ~=~ e^{At} \xb , 
		\hstm \hstm 
		\mbox{since}~~& ~
		\dot{x}(t) ~=~ \frac{d}{dt} e^{At} \xb ~=~ A ~ e^{At} \xb ~=~ A ~x(t) , 			\\
		~\mbox{and}~~ & ~ 
		\left. x(0) ~=~ e^{At} \xb \right|_{t=0} ~=~  e^{0} \xb ~=~ I \xb ~=~ \xb. 
	 \end{aligned} 
	 \label{exp_At_sol.eq}
	\ee
	The solution  of~\req{ss_PB} with a non-zero forcing function $w$ is given by the 
	``variations-of-constants'' formula\footnote{This is also known as the Cauchy formula.}  
	\be
		x(t) ~=~ e^{At} \xb ~+~ \smint{0}{t} e^{A(t-\tau)} ~w(\tau) ~d\tau. 
	  \label{VOC_LTI.eq} 
	\ee
	The fact that this formula gives $x(t)$ that satisfies the differential equation~\req{ss_PB} 
	 can be directly verified by differentiation. 
	
	The development described above, while quick and expedient, is unsatisfactory. Although it might be 
	easy to guess the definition of the matrix exponential, and the solution~\req{exp_At_sol} as 
	a generalization of the well known scalar case, it is difficult to see how this might generalize to the 
	linear time-varying or the nonlinear cases. Again,  from a pedagogical point of view, being told what the answer 
	to a problem 
	is, and your role is simply to verify that it is indeed the answer is not helpful in gaining insight into 
	how more general situations might be addressed. For example, if you have not seen the 
	formula~\req{VOC_LTI} before, 
	it probably seems to ``come out of thin air''. It is easy to verify, but where did it come from? What would a
	similar formula be in the time-varying case, or if the state $x$ is a matrix rather than a vector?

	A more satisfactory development is to see the 
	answer emerge naturally from basic, familiar  principles that are applicable to a large variety of 
	problem settings. 
	At the expense of a little bit of abstraction, we can have a better, more contextual understanding of 
	the subject. 
	This is the approach we will follow in this paper which is 
	organized around the following central idea.
	 The differential
	equations~\req{SS_non} are rewritten as integral equations, which   can then be thought of as equations 
	in an abstract  function space involving the {\em Volterra operator of forward integration}. This operator has 
	very special properties which we investigate. The various series expressions and iterative algorithms for 
	solutions follow from Neumann series involving this operator. In particular, the matrix exponential, the Peano-Baker 
	series, the variations-of-constants formula, and the Picard iteration are all specific manifestations of this abstract
	Neumann series. They all emerge naturally from applying the Neumann series without having to guess the answer. 
	Furthermore, the convergence properties of all these series and iterations 
	 follow from an {\em asymptotic nilpotence} property of the Volterra operator. 
	This gives a unified view of all the various results in this area. 
	
	This presentation is organized as follows. 
	\begin{description} 
		\item[Section~\ref{basic.sec}:]  Introduces the basic properties of flow maps, and their special forms when 
				the dynamics are time-invariant or linear respectively. These properties follow from 
				the basic assumptions of existence and uniqueness. 
		\item[Section~\ref{abstract.sec}:] Recasts the solution of state-space models in the linear case as a linear 
			algebra problem in function space. The Volterra (forward) integration operator is
			introduced as analogous to strictly lower-triangular matrices. Such matrices are nilpotent, 
			and the Volterra operator is shown to be ``asymptotically nilpotent''. Convergence  of series and 
			iterations with the Voiterra operator then follow from this latter property. The {\em kernel representation 
			of linear operators} is introduced here as the main tool to understand these properties. 
		\item[Section~\ref{Form_STM.sec}:] Shows how the matrix exponential and the Peano-Baker series are special instances
			of the Neuman series. 
		\item[Section~\ref{inputs.sec}:] Considers systems with inputs in the linear case. The well-known 
			``variations-of-constants'' (Cauchy) formula is derived in three different ways, one of which 
			is again as a consequence of the Neumann series. Readers not interested in systems with 
			inputs can skip this section. 
		\item[Section~\ref{Picard.sec}:] Considers general non-linear systems. The Neumann series here becomes the 
			Picard iteration. The proof of convergence follow from fixed point theorems. The 
			standard contraction mapping theorem is used to show local existence and uniqueness. 
			A tighter fixed point theorem that uses the asymptotic nilpotence 
			of the Volterra operator is used to show global  existence and uniqueness. 
	\end{description}

	\section{Basic Properties} 									\label{basic.sec}
	
	We consider first systems without inputs (systems ({\it a}), ({\it c}) and ({\it e}) in~\req{SS_non}). 
	We will give various conditions for existence and 
	uniqueness of solutions later on  in Section~\ref{Picard.sec}. 
	For now however, we will make the following standing assumption. 
	\begin{definition} 
		A system of the form~\req{SS_non} without input is said to be {\em well posed} over a time 
		interval $(a,b)\subseteq\R$ if for any initial time $\tb\in(a,b)$, and 
		any  initial condition $x(\tb)\in\R^n$, there exists a unique solution of the differential equation 
		over $(a,b)$. $a$ or $b$ may be $-\infty$ or $\infty$ respectively, and intervals can also be half 
		open (i.e. $[a,b)$ or $(a,b]$). 
	\end{definition}
%
	 It should be noted that a requirement of existence and uniqueness of solutions 
	is a natural one when differential equations are used as mathematical models of physical phenomena. 
	If solutions are non-unique, then the model must be missing some feature of the physical phenomenon, 
	and likewise if solutions do not exist for some initial conditions.  Further commentary along these 
	lines is included in Section~\ref{commentary.sec} after the conditions for existence and uniqueness are established.

	 \subsection{The Flow Map}

	 Consider now any of the systems  ({\it a}), ({\it c}) and ({\it e}) in~\req{SS_non}, and 
	 assume well-posedness over $[0,T)$ for some $T$. 
	 The existence and uniqueness of solutions assumption implies 
	 that for each $t,\tb\in[0,T)$ there is a well-defined mapping $\Phi_{t,\tb}:\R^n\rightarrow\R^n$ such that 
	 \be
	 	x(t) ~=~ \Phi_{t,\tb}\big( x(\tb) \big) . 
	  \label{flow_map.eq} 
	 \ee
	 We don't know this mapping explicitly. This mapping is simply the statement that $x(t)$ is the solution 
	 of the differential equation at time $t$ given the initial condition $x(\tb)$ at time $\tb$. Since solutions
	  exist and are unique by assumption, this is a well defined mapping. More precisely, 
	  $\Phi:=\lcb \Phi_{t,\tb}, ~t,\tb\in[0,\infty) \rcb$ is a two-parameter family 
	  of mappings on $\R^n$. We refer to $\Phi$ as the 
	  {\em flow map} of the dynamical system.  

\begin{figure}[t]
		\centering
		\begin{subfigure}[b]{0.5\textwidth} 
			\centering 
			\includegraphics[width=\textwidth]{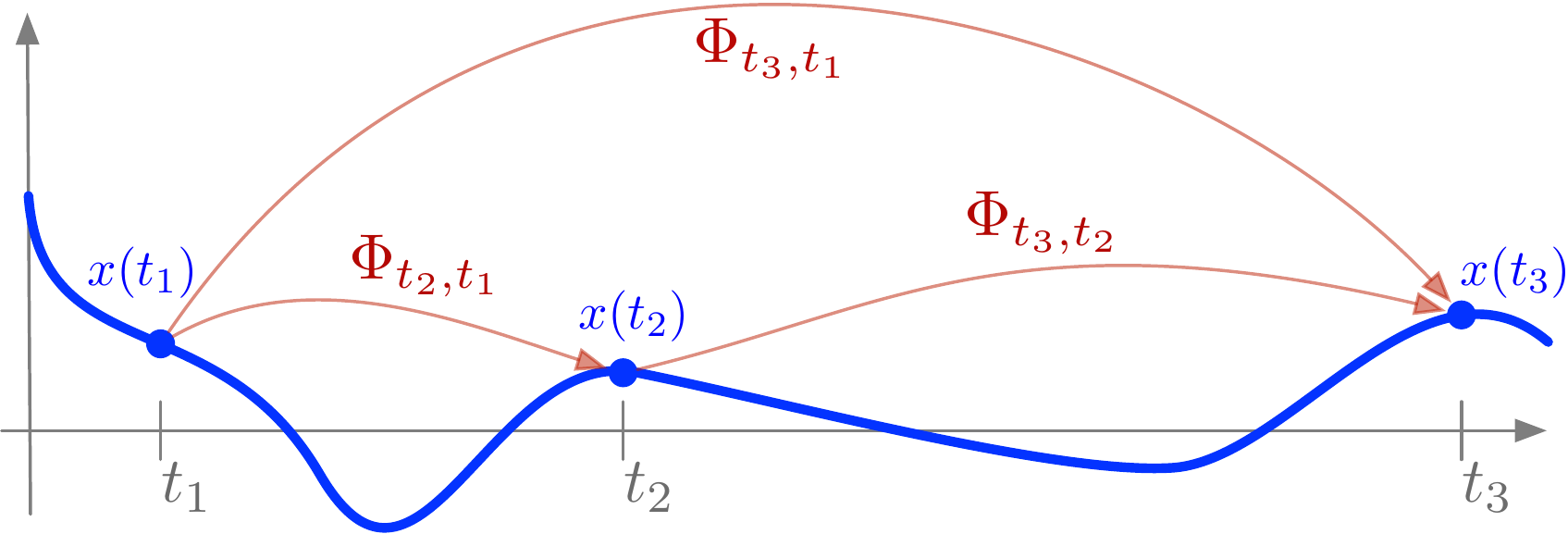}
			
			\mysubcaption{The semigroup property 
			$\Phi_{t_3,t_1}=\Phi_{t_3,t_2} \circ \Phi_{t_2,t_1}$ 
			is equivalent to saying that finding the solution at time $t_3$ given an initial 
			condition at time $t_1$ is equivalent to solving in two steps. First, find the solution $x(t_2)$ 
			at some intermediate time $t_2$ from the initial condition at $t_1$, then find the 
			solution at time $t_3$ from $x(t_2)$ regarded as an
			 initial condition at $t_2$. The answer should be the same 
			as the going directly from $t_1$ to $t_3$. 			
				} 
			\label{semigroup_SS.fig} 
		\end{subfigure} 
		\quad
		\begin{subfigure}[b]{0.45\textwidth} 
			\centering 
			\includegraphics[width=.8\textwidth]{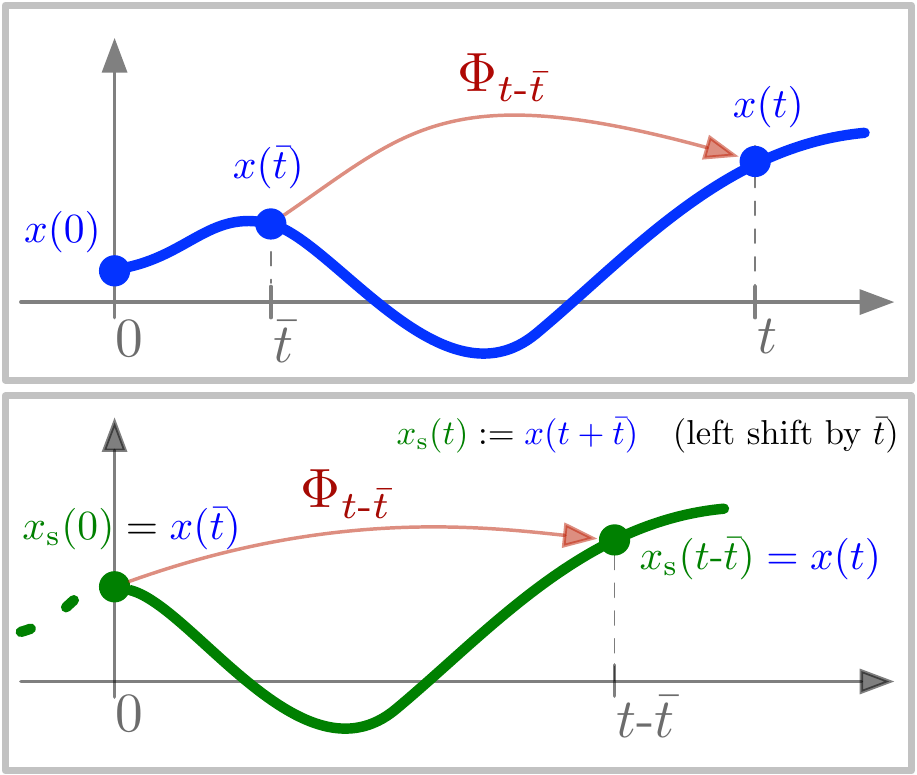}
			
			\mysubcaption{The time invariance property  implies that the solution $x(t)$ given 
				an initial condition $x(\tb)$ is the same as solving a time-shifted problem (depicted in green), 
				i.e. 	solving for  $x_{\rm s}(t-\tb)$ from an initial condition $x_{\rm s}(0)=x(\tb)$.
				 } 
			\label{TI_Ht_SS.fig}
		\end{subfigure} 
		
		\mycaption{Well posed-ness (existence and uniqueness of solutions) 
			 of the system~\req{SS_non}  implies that the solution $x(t_2)$ at any time $t_2$ from an initial 
			 conditions $x(t_1)$ at time $t_1$ is given by a well-defined, two-parameter  family of mappings 
			 $\Phi_{t_2,t_1}:\R^n\rightarrow\R^n$.
			Well posedness also implies the semigroup property depicted in (a). When the system has 
			constant (in time) coefficients, then the time-invariance (more precisely time-shift equivariance) 
			 property holds as depicted in (b), and the flow map is given by a single-parameter family 
			 $\Phi_{t_2,t_1}=\Phi_{t_2\sm t_1,0}$. 
					} 
		\label{TI_semigroup_SS.fig}
	\end{figure}

	The flow map's dependence on the parameters  has  properties that follow 
	immediately from its definition. First, at any 
	$t\in[0,T)$ 
	\[
		\Phi_{t,t} ~=~ I, 
	\]
	where $I$ is the identity map. This follows since $\Phi_{t,t}$ maps an initial condition at $t$ to the 
	solution at $t$, i.e. it maps each vector to itself.  
	Second,  consider three time instants $t_1,t_2,t_3\in[0,T)$. 
	If we solve the equation from $t_1$ to $t_3$ starting from $x(t_1)$, then the solution $x(t_3)$ 
	must be the same as what is obtained by solving the equation from $t_1$ to $t_2$ and then 
	again from $t_2$ to $t_3$, with the latter  starting from $x(t_2)$ as an initial condition
	(see Figure~\ref{semigroup_SS.fig} for an illustration). 	
	In other words 
	\[
		x(t_3) ~=~ \Phi_{t_3,t_1} \big( x(t_1) \big) 
			~=~ 	 \Phi_{t_3,t_2} \big( x(t_2) \big) 
			~=~ 	 \Phi_{t_3,t_2} \lb   \Phi_{t_2,t_1} \big( x(t_1) \big) \rom \rb
			~=~ \lbb \Phi_{t_3,t_2} \circ   \Phi_{t_2,t_1} \rom \rbb \big( x(t_1) \big) , 
	\]
	where the symbol $\circ$ denotes function composition. Since this has to hold for all possible initial 
	conditions,  we have  equality of the mappings 
	\be
		 \Phi_{t_3,t_1} ~=~ \Phi_{t_3,t_2} \circ   \Phi_{t_2,t_1} 
	 \label{semigroup_TV.eq}
	\ee
	for all $t_1,t_2,t_3\in[0,T)$. This property is called the {\em semigroup property}, although this name 
	is better suited for the time-invariant case which we discuss next.

	\subsection{Time Invariance}

	The concept of time invariance requires existence and uniqueness of solutions over semi-infinite time
	 intervals. Without loss of generality, we will therefore assume well-posedness of the following system 
	 over the entire half line $[0,\infty)$
	\be
		\xd(t) ~=~ A \big( x(t) \big) , 
		\hstm \hstm 
		t\in[0,\infty) , 
	  \label{xdF_TI.eq}
	\ee
	where the vector field $A$ is constant in time.
	The ``dynamics'' of
	 this system (i.e. the relation between $\xd$ and $x$ at each time) are independent of $t$. Suppose 
	 that $x(.)$ is the solution from the initial condition $x(0)=\xb$. Define the {\em left-shift} of $x$ by 
	 \[
	 	x_{\rm s} (t) ~:=~ x(t+\tb), 
		\hstm \hstm 
		t\in[0,\infty) , 
	 \]
	 where $\tb$ is some fixed number. Note that the initial condition for $x_{\rm s}(0)$ is $x(\tb)$. 
	 Now  if we solve the equation from the initial condition $x_{\rm s}(0)=x(\tb)$, the solution will simply 
	 be the portion of the original trajectory  over $[\tb,\infty)$, which is the same as $x_\rms$ over $[0,\infty)$
	 \[ 
	 	\left. 
	 	\arraycolsep=2pt
	 	\begin{array}{rcl} 
	 		\xd_{\rm s}(t) &=& \xd(t+\tb) 					\\
			A\big( x_\rms(t) \big) &=&  A\big( x(t+\tb) \big)	\\
		\end{array} 		
		\right\} 
		\hsom \Rightarrow \hsom 
		 \xd(t+\tb) = A\big( x(t+\tb) \big)
		\hsom \Rightarrow \hsom 
		\left\{
	 	\arraycolsep=2pt
	 	\begin{array}{rcl} 
	 		\xd_{\rm s}(t) 	&=& A\big( x_{\rm s}(t) \big) 			\\
			x_{\rm s} (0) 	&=& x(\tb)						\\
		\end{array} 		
		\right.  .
	 \]
	 Thus the shifted function $x_\rms$ satisfies the differential equation with its initial condition as 
	 the vector $x(\tb)$. Figure~\ref{TI_Ht_SS.fig} illustrates this property, which is called time invariance
	 (or more precisely time-shift equivariance), for which we give a formal definition. 
	 \begin{definition} 
	 	Consider a state space system over $[0,\infty)$. The system is called {\em time invariant} 
		 if for each solution trajectory 
		$\lcb x(t), ~t\in[0,\infty) \rcb$, and each $\tb\geq 0$,  the left-shifted trajectory 
		$x_{\rm s}(t):=x(t+\tb)$ is a solution starting from the initial condition $x_\rms(0)=x(\tb)$. 
	 \end{definition} 
	 
	 The previous calculations show that a system of the form~\req{xdF_TI} where the vector field 
	 $F$ is independent of $t$ satisfies the definition above for time invariance. 
	 Note that the key 
	 to this fact is  $\xd(t+\tb) =A\big( x(t+\tb) \big)$, which would not be true if the vector field $A$ 
	 depended on time since in general $A\big( x(t+\tb),t+\tb \big) \neq A\big( x(t+\tb),t \big)$.

	 Time invariance has an important implication for the dependence of the flow  $\Phi_{t_2,t_1}$ 
	 on the parameters $t_1$ and $t_2$. 
	 Let $x$ be any trajectory, and let $x_\rms(t):=x(t+t_1)$ be its shifted version.	 
	  In terms of the flow 
	 maps, we have the relations 
	 \begin{align*} 
	 	x_\rms(t) ~:=~ x(t+t_1)		
		\hstm &\Rightarrow \hstm 
		x_\rms(0) = x(t_1) 
		~\mbox{and}~ 
		x_\rms(t_2-t_1) = x(t_2) 		,		\\ 
	 	\left. 
	 	\arraycolsep=2pt
	 	\begin{array}{rcl} 
	 		x(t_2) 	&=& 		\Phi_{t_2,t_1} \big( x(t_1) \big)			\\
			x_\rms(t_2-t_1)  &=&  \Phi_{t_2\sm t_1,0} \big( x_\rms(0) \big)	\\	
								&=&	 \Phi_{t_2\sm t_1,0} \big( x(t_1)  \big)
		\end{array} 		
		\right\} 
		&	\Rightarrow \hstm 
		\Phi_{t_2,t_1} \big( x(t_1) \big)	 = \Phi_{t_2\sm t_1,0} \big( x(t_1)  \big)
	 \end{align*} 
	Since this last statement must hold  
	 for all possible vectors $x(t_1)$, we conclude that as maps 
	 \[
	 	\Phi_{t_2,t_1} ~=~ \Phi_{t_2\sm t_1,0}. 
	 \]
	 Thus the two-parameter family of flow maps is fully determined by the one parameter family 
	 $\lcb \Phi_{t,0}, ~t\in[0,\infty) \rcb$. With a slight abuse of notation, we relabel this family as $\Phi_t$
	 and state this conclusion formally. 
	 \begin{lemma} 
	 	Assume the time-invariant dynamical system~\req{xdF_TI} is well posed over $[0,\infty)$. Then 
		there exists a one parameter family of maps $\Phi_t:\R^n \rightarrow \R^n$, $t\in[0,\infty)$
		such that for any trajectory, and any times $t_2\geq t_1 \geq 0$  
		\[
			x(t_2) ~=~ \Phi_{t_2\sm t_1} \blb x(t_1) \brb.
		\] 
		In particular, $x(t) ~=~ \Phi_{t} \blb x(0) \brb$. 
	 \end{lemma}
	 
	 Thus for time-invariant systems, the flow map between the state at time $t_1$ and the state at time 
	 $t_2$ {\em depends only on the time difference} $t_2-t_1$, and not on the starting or ending times $t_1$ 
	 and $t_2$. Now recall the so-called  semigroup property~\label{semigroup_TV}, which when combined 
	 with time invariance says 
	\begin{align*} 
	 		 \Phi_{t_3,t_1} = \Phi_{t_3,t_2} \circ   \Phi_{t_2,t_1} 
			 \hstm &\Rightarrow \hstm 
	 		 \Phi_{t_3\sm t_1,0} = \Phi_{t_3\sm t_2,0} \circ   \Phi_{t_2\sm t_1,0} 		\\
			 \hstm &\Rightarrow \hstm 
	 		 \Phi_{t_3\sm t_1} = \Phi_{t_3\sm t_2} \circ   \Phi_{t_2\sm t_1} , 
	\end{align*} 
	where we repeated the abuse of notation in the last statement. If we now relabel $t_2-t_1=:\tau_1$ 
	and $t_3-t_2=:\tau_2$, then $t_3-t_1=\tau_2+\tau_1$ and we can finally state that for  
	time-invariant systems, the one-parameter family of flow maps satisfies the property 
	\be
		\Phi_{\tau_2} \circ \Phi_{\tau_1} ~=~ \Phi_{\tau_2+\tau_1} , 
		\hstm \hstm 
		\tau_1,\tau_2\in [0,\infty). 
	 \label{semigroup_TI.eq}
	\ee
	This can now be legitimately called the {\em semigroup property}. There is a one-to-one correspondence between
	the set $[0,\infty)$ (which is a semigroup under addition) and the family of maps 
	$\Phi := \lcb \Phi_t, ~t\in[0,\infty) \rcb$. The property~\req{semigroup_TI} 
	is a statement of semigroup isomorphism. $\Phi$ is a semigroup with the operation of function 
	composition, which is isomorphic to $[0,\infty)$, a semigroup under addition.

	\subsection{The State Transition Matrix} 								\label{STM.sec}

	In the case of linear systems, the flow map $\Phi_{t_2,t_1}$ can be shown to be 
	a {\em linear mapping} on $\R^n$, and is 
	therefore represented by a {\em matrix-valued} function $\Phi(t_2,t_1)$ of two parameteres, 
	which is naturally called the {\em state transition matrix}. 
	The fact that the flow map is linear is easy to show 
	 without actually ``solving'' the equation
	 as follows. Consider a linear time-varying system with no input, and
	 two solutions corresponding to two initial conditions 
	 \begin{align*} 
	 	\xd_1(t)  ~&=~ A(t) ~x_1(t) , 
				\hstm \hstm x_1(\tb) ~=~ \xb_1, 			\\
	 	\xd_2(t)  ~&=~ A(t) ~x_2(t) , 
				\hstm \hstm x_2(\tb) ~=~ \xb_2. 
	 \end{align*} 
	 We can immediately verify that the solution due to a linear combination $x(\tb)=\alpha \xb_1 + \beta \xb_2$ of the 
	 initial conditions is the same linear combination $x(t) = \alpha x_1(t)+\beta x_2(t)$  of the individual solutions. Indeed   
	 \begin{align*} 
	 	\frac{d}{dt} \big(\alpha  x_1(t) + \beta x_2(t) \big) 
		~&=~ 
		 \alpha~  \xd_1(t) + \beta~ \xd_2(t)
		~=~ 
		\alpha  ~A(t) ~x_1(t) + \beta ~A(t) ~x_2(t) 							\\
		&=~ 
		A(t) ~\big( \alpha   ~x_1(t) + \beta ~x_2(t) \big) 
	 \end{align*} 
	 Thus $x(t) = \alpha x_1(t)+\beta x_2(t)$ satisfies the differential equation as well as the initial condition. Note that the 
	 only property used above is the linearity of differentiation and the linearity of the right hand side of the differential 
	 equation.

	 Recall that the flow map $\Phi_{t,\tb}$ maps initial conditions at $\tb$ to solutions at $t$. Since we have 
	 established that this map is linear, and on $\R^n$ general linear maps are  represented by matrices, 
	 then there must exist a matrix-valued 
	 function of time $\Phi(t,\tb)$ 
	 such that 
	 \begin{align} 
	 	x(t) ~&=~\Phi_{t,\tb} \blb x(\tb) \brb , 
				&		 \Phi_{\tb,\tb} ~&=~ I \hstm \mbox{(the identity mapping)} 		\nonumber		\\
	 	 \Rightarrow	\hstm
		 x(t) ~&=~ \Phi(t,\tb) ~x(\tb) 
		  		& 		\Phi(\tb,\tb) ~&=~I \hstm \mbox{(the identity matrix)} 
	  \label{STM_first.eq}
	 \end{align} 
	 This matrix-valued function of two time parameters $\Phi(.,.)$ is naturally 
	 called the {\em state transition matrix}. It inherits properties of the flow map when 
	 specialized to linear maps. For example, $\Phi_{t,t}$ is the identity mapping for any $t$, 
	 and since this linear map is represented by the matrix $\Phi(t,t)$, this must be the 
	 identity matrix. The semigroup property is also inherited, and in this case composition 
	 of maps becomes matrix multiplication. We now state these properties formally. 
%
%
%
%
	 \begin{lemma} 												\label{STM.lemma}
	 	Consider the linear time-varying  system 
			\be
	 				\xd(t) ~=~ A(t) ~x(t) , 
					\hstm \hstm x(\tb)=\xb, 
			  \label{LTV_STM_form.eq}
			\ee
			 and assume it is well posed over $\R$.  
			 Then
                    			 there exists a matrix-valued function, called the 
                    			 {\em state transition matrix} $\Phi(.,.)$,   such that for any two times $t_1,t_2\in\R$ 
                    			\be
                    				x(t_2) ~=~ \Phi(t_2,t_1) ~x(t_1). 
					  \label{STM_sol.eq}
                    			\ee
					This state transition matrix has the {\em semigroup property} where for any $t_1,t_2,t_3\in\R$ 
                    			\be
                    				\Phi(t_3,t_1)  ~=~  \Phi(t_3,t_2) ~\Phi(t_2,t_1) . 
					  \label{STM_semigroup.eq}
                    			\ee
					In particular, it is always non-singular and inverses are given by 
					$
						\Phi^{- 1}(t_2,t_1) = \Phi(t_1,t_2) . 
					$
					Furthermore, it  satisfies the matrix differential equation 
					\be
						\frac{d}{dt} \Phi(t,\tb) ~=~ A(t) ~ \Phi(t,\tb) , 
						\hstm \hstm 
						\Phi(\tb,\tb) ~=~ I, 
						\hstm \mbox{for any}~\tb\in\R. 
					  \label{STM_diffeq.eq}
					\ee
	 \end{lemma} 
	 
	 In the last section, we will show that this system is well posed on $\R$ under the reasonable assumption 
	 that the function $A(.)$ is 
	 bounded on bounded intervals. Well-posedness over all of $\R$
	  implies that the system can be solved forward or backwards in time 
	 from any $\tb\in\R$. 
	 
	 The invertibility of the state transition matrix follows from the semigroup property since we can always 
	 expresses the inverse as
	 \[
	 	\Phi(t_1,t_2) ~\Phi(t_2,t_1) ~=~ \Phi(t_1,t_1) ~=~ I. 
	 \]
	 In other words, solving from an initial condition $x(t_1)$ at time $t_1$ to the response $x(t_2)$ 
	 at time $t_2$, and then solving backwards from $x(t_2)$ to the state at time $t_1$ should given 
	 the original $x(t_1)$.  
	 Finally,  the differential equation~\req{STM_diffeq} 
	can be  verified from the formula~\req{STM_first} together with the original 
	differential equation~\req{LTV_STM_form} 
	\[
		\left. \arraycolsep=2pt
		\begin{array}{rcl} 
			\xd(t) &=&  A(t) ~x(t) 		\\ 
			x(t) &=& \Phi(t,\tb) ~x(\tb)
		\end{array} 
		\right\} 
		\hstm \Rightarrow \hstm 
		\frac{d}{dt} \Phi(t,\tb) ~x(\tb) ~=~ A(t)~ \Phi(t,\tb) ~x(\tb) .
	\]
	Since the last equality holds for any initial condition vector $x(\tb)$, then the matrix equation~\req{STM_diffeq}
	must hold. 
	
	Another new ingredient here over the nonlinear case is the differential equation~\req{STM_diffeq} for the 
	state transition matrix. We note that it is possible to derive an analogous equation for the nonlinear
	flow map $\Phi$ (see Exercise~\ref{Flow_PDE.ex}), but it is a partial differential equation, and not particularly 
	useful for computations due to the ``curse of dimensionality''.


	It is important to keep in mind that Lemma~\ref{STM.lemma} does not ``solve'' the system~\req{LTV_STM_form}
	in any concrete sense. It just states the properties of the solution. The ``solution formula''~\req{STM_sol} 
	is simply an 
	expression of the linearity of the problem. The semigroup property is a consequence of the uniqueness 
	of solutions. The differential equation~\req{STM_diffeq} for $\Phi$ is actually more complicated to solve 
	than the original system when given a specific initial condition. The latter is a vector differential equation, while the 
	former is a matrix differential equation. However, as we will now demonstrate, if we solve $n$ vector differential 
	equations from properly chosen initial conditions, we can find the state transition matrix. 
	
	Let $\lcb v_k \rcb_{k=1}^n$ be any basis of $\R^n$.
	Suppose that we numerically solve the $n$, vector differential 
	equations 
	\be
		\xd_k(t) ~=~ A(t) ~ x_k(t) , 
		\hstm\hstm 
		x_k(\tb) ~=~ v_k , 
		\hstm 
		k=1,\ldots,n, 
	  \label{STM_n_vec_sol.eq}
	\ee
	each with $v_k$ as its initial condition. 
	Those $n$ solutions can be used to obtain the solution for other initial conditions. 
	Expand any initial condition  $x(\tb)=\xb$ using the basis, and write 
	this in matrix-vector form as 
	\[
		\xb ~=~ \sum_{k=1}^n \alpha_k v_k
		\hstm \Leftrightarrow \hstm 
			\bbm \\ \xb \\ ~\ebm 
			~=~  
					\left[\arraycolsep=2pt
						 \begin{array}{c:c:c} 	
							v_1 & \cdots & v_n 		\rule{0em}{1.8em} \\
								\vphantom{x} &  &  
						\end{array} \right] 			
				\bbm \alpha_1 \\ : \\ \alpha_n \ebm 
			~=:~ V ~\alpha 
	\]	
	Since the 
	mapping from $x(\tb)$ to $x(t)$ is linear, then 
	\[
		x(t) ~=~ \sum_{k=1}^n \alpha_k x_k(t)  
			~=~ 
			 \left[\arraycolsep=2pt
						 \begin{array}{c:c:c} 	
							x_1(t) & \cdots & x_n(t) 		\rule{0em}{1.8em} \\
								\vphantom{x} &  &  
						\end{array} \right] 	
				\bbm \alpha_1 \\ : \\ \alpha_n \ebm 
		~=~ \Phi(t,\tb) ~\xb ~=~  \Phi(t,\tb) ~V~\alpha . 
	\]
	Since this must hold for every initial conditions (i.e. for every vector $\alpha$ of coefficients), then 
	\be
			\Phi(t,\tb) ~=~  \left[\arraycolsep=2pt
						 \begin{array}{c:c:c} 	
							x_1(t) & \cdots & x_n(t) 		\rule{0em}{1.8em} \\
								\vphantom{x} &  &  
						\end{array} \right] 
					\left[\arraycolsep=2pt
						 \begin{array}{c:c:c} 	
							v_1 & \cdots & v_n 		\rule{0em}{1.8em} \\
								\vphantom{x} &  &  
						\end{array} \right]^{-1}  			
	  \label{STM_vk_basis.eq}
	\ee
%
%
	In other words, given a state dimension of $n$, if we choose a basis $\lcb v_k\rcb_{k=1}^n$
	and  solve the $n$ vector differential 
	equations~\req{STM_n_vec_sol}, then the state transition matrix is obtained from those 
	$n$ vector functions of time by~\req{STM_vk_basis}. 
	Note that this formula satisfies $\Phi(\tb,\tb)=VV^{-1}=I$ as required. 
	A particularly simple choice of basis 
	is the canonical basis $\lcb e_k\rcb_{k=1}^n$. The matrix $V$ in this case is the identity, 
	and according to~\req{STM_vk_basis} the state transition matrix would simply be made 
	from the $n$ solutions $\lcb x_k(t) \rcb_{k=1}^n$ as its columns.

	\section{A Linear Algebra Problem in Function Space} 					\label{abstract.sec}

	The key to the solution formulas for the system~\req{ss_PB} is a slight abstraction where we think 
	of the system as a linear algebra problem but in function space. 
	From this point of view, it is at first just as easy to do the time-varying case, which is a system of 
	the form
	\be
		\xd(t) ~=~ A(t) ~x(t) ~+~ w(t), 
		\hstm \hstm x(0)=\xb
	 \label{ss_PB_TV.eq} 
	\ee
	This equation is equivalent to an integral equation which we obtain by  
	 Integrating both sides of~\req{ss_PB_TV}  
	\be
		x(t) ~-~ x(0) ~=~ \smint{0}{t} A(\tau) ~x(\tau) ~d\tau ~+~ \smint{0}{t} w(\tau) ~ d\tau. 
	  \label{ss_PB_int.eq}
	\ee
	To express this equation as a linear algebra problem, fix a time horizon $[0,T]$, 
	and define the {\em Volterra integration operator}, which 
	we denote by the  symbol $\Volt$ 
	\be
		\big( \Volt g \big) (t)  ~:=~ \smint{0}{t} g(\tau) ~d\tau. 
	  \label{int_op_def.eq}
	\ee
	This operator is well-defined on the function space $\LP{1}_n[0,T]$. Define also the operator 
	\[
		\big( \cA g \big) (t) ~:=~ A(t) ~g(t) 
	\]
	of point-wise (in time) multiplication by $A(.)$. Finally define the operator $\heavi:\R^n\rightarrow \LP{1}_n[0,T]$, 
	which takes vectors $\xb\in\R^n$  to constant functions of time by 
	\[
		\big( \heavi \xb \big) (t) ~:=~ \heavi(t) ~\xb, 
		\hstm\hstm t\in[0,T], 
	\]
	where $\heavi(.)$ is the unit step (Heaviside) function
	\[
		\heavi(t) ~:=~ \left\{ \begin{array}{lcl} 
						1	& & t\geq 0 		\\ 
						0 	& & t<0
						\end{array} \right. 	.
	\]
	 Note the slight abuse of notation where we use the same 
	symbol to denote a function $\heavi(.)$ of time, as well as this operator. 
	
	With the above definitions, the integral equation~\req{ss_PB_int} can now be written as the abstract equation 
	\[
		x ~=~ \Volt \cA ~x ~+~ \Volt ~w ~+~ \heavi ~\xb, 
	\]
	where $\Volt \cA$  is the composition of the action of the operator $\cA$ (first) with the operator $\Volt$ (second). 
	Since $w$ and $\xb$ are usually given, we rewrite this equation so as to solve for $x$ in terms of the given 
	quantities by 
	\be
		\big( I  - \Volt \cA \big)  ~x ~= ~ \Volt w + \heavi \xb. 
	  \label{SS_sol_RHS.eq}
	\ee
	The right hand side $( \Volt w + \heavi \xb)$ and $x$ are functions over $[0,T]$, and $\big( I - \Volt \cA \big)$ 
	is a linear operator on such 
	functions. If this operator is invertible, then  the solution is
	\be
		\boxed{~\rom
		x ~=~ \big( I  - \Volt \cA \big)^{-1}   \big(  \Volt w + \heavi \xb \big) . 
		~}
	  \label{SS_sol_Volt.eq}
	\ee
	Thus we need to understand the operator $\big( I  - \Volt \cA \big)^{-1}$ and its properties. 
	The key is the familiar  Neumann series 
	\be	\textstyle
		 \big( I  - \Volt \cA \big)^{-1} 
		 ~=~ I ~+~ \Volt \cA ~+~ \big( \Volt \cA \big)^2 ~+~ \big( \Volt \cA \big)^3 ~+~ \cdots  
		 ~=~ \sum_{n=0}^\infty \big( \Volt \cA\big)^n . 
	  \label{Neumann_PB.eq}
	\ee
	
	The Neumann series has  an interpretation as an iterative algorithm generally known as a {\em fixed point iteration}. 
	Denote the right hand side of~\req{SS_sol_RHS} by $g$. The solution in terms of the Neumann series is then 
	\begin{align}	\textstyle
		x ~=~ 
		 \big( I  - \Volt \cA \big)^{-1} ~g
		 ~&=~ g~+~ \Volt \cA~g ~+~ \big( \Volt \cA \big)^2 ~g  ~+~ \big( \Volt \cA \big)^3  ~g ~+~ \cdots 
		 															 	  \label{Neu_series_sol.eq}  \\
		&=~ g ~+~ \lb \Volt \cA ~\lb g ~+~ \Volt \cA~ \lb g ~+~ \Volt \cA ~\lb g ~+~ \cdots ~\rb \rb \rb \rb 	\nonumber
	\end{align} 
	This infinite series can be rewritten as the iterative algorithm
	\be
            	\begin{aligned} 
            		x_0 ~&=~ g , 		\\ 
            		x_{k+1} ~&=~ g ~+~ \Volt \cA ~ x_k .
            	\end{aligned} 
	  \label{picard_it.eq}
	\ee
	Thus $x_k$ is the $k$'th partial sum of the series~\req{Neu_series_sol}. 
	If this series converges, then the limit $x:=\lim_{k\rightarrow\infty} x_{k} = \lim_{k\rightarrow\infty} x_{k+1}$ satisfies 
	\[
		x ~=~ g ~+~ \Volt\cA ~x 
		\hstm \Leftrightarrow \hstm 
		\big( I - \Volt\cA\big) ~x ~=~ g, 
	\] 
	and thus the limit of the iteration~\req{picard_it} is indeed a solution of the original problem. We will study the convergence 
	properties of this iteration, which also is applicable to nonlinear problems under certain conditions. For now, we consider 
	only linear problems.

	It turns out that the Volterra  integration operator $\Volt$ has a special property that guarantees the convergence 
	of the Neumann series~\req{Neumann_PB}  under 
	very mild conditions. In addition, this formula will lead naturally to the matrix exponential when $A$ is constant, and to the 
	so-called Peano-Baker series in the time-varying case. First, we need to establish some important properties of $\Volt$.

	The Volterra integration operator $\Volt$ is analogous to a strictly lower triangular matrix with entries of $1$ below 
	the diagonal. This analogy is important to understand properties of this operator, and it is best done using the 
	so-called {\em kernel representation} of linear operators which we now describe. 
	
	\subsection{The Kernel Representation of Linear Operators}

		Let $A$ be an  $n\times n$ matrix with the $ij$'th entry denoted by $A_{ij}$. A matrix represents a linear 
		operator on vectors by the 
		matrix vector product $v=Au$
		\be
			v~=~ A~u
			\hstm \Leftrightarrow \hstm 
			v_i ~=~ \sum_{j=1}^n A_{ij} u_j 		.
		  \label{m_vect_mult.eq} 
		\ee
		Now let $\cI=(a,b) \subseteq\R$ be any interval, and let $A(.,.)$ be a real-valued 
		function\footnote{In the case when $u$ and $v$ are vector-valued functions, then $A(.,.)$ would be 
		a matrix-valued function. We suppress this distinction in our notation, which is 
		equally applicable to either situation. } 
		of two variables 
		from that interval $A:\cI\times\cI\rightarrow\R$. Such a function defines a linear operator on single-variable
		 functions over $\cI$ 
		in an analogous manner to~\req{m_vect_mult} by 
            \be
            	v ~=~ A ~u 
		\hstm \Leftrightarrow \hstm 
            	v(x) ~=~ \int_{\cI} A(x,\xi) ~u(\xi) ~d\xi , 
			\hstm x\in\cI, 
              \label{a_kernel.eq}
            \ee
            where the integration variable $\xi$ plays the same role as the column index $j$ over which the summation 
            in~\req{m_vect_mult} is performed. 
            The operation in~\req{a_kernel} is a linear 
            operator $A: u \mapsto v$, and note the slight abuse of notation where we use the same symbol $A$ to denote the 
            operator, as well as the function $A(.,.)$ of two variables. 
            The  function $A(.,.)$ is called the {\em kernel function} of the operator $A$, and the 
            formula~\req{a_kernel} is called the 
            {\em kernel representation}\footnote{The reader should be careful not to confuse this with the null space of the
            operator, which is sometimes referred to as the kernel of the operator. The two concepts are unrelated.} 
            of $A$.

            
            \begin{figure}[t]
            	\centering
            		\includegraphics[width=0.85\textwidth]{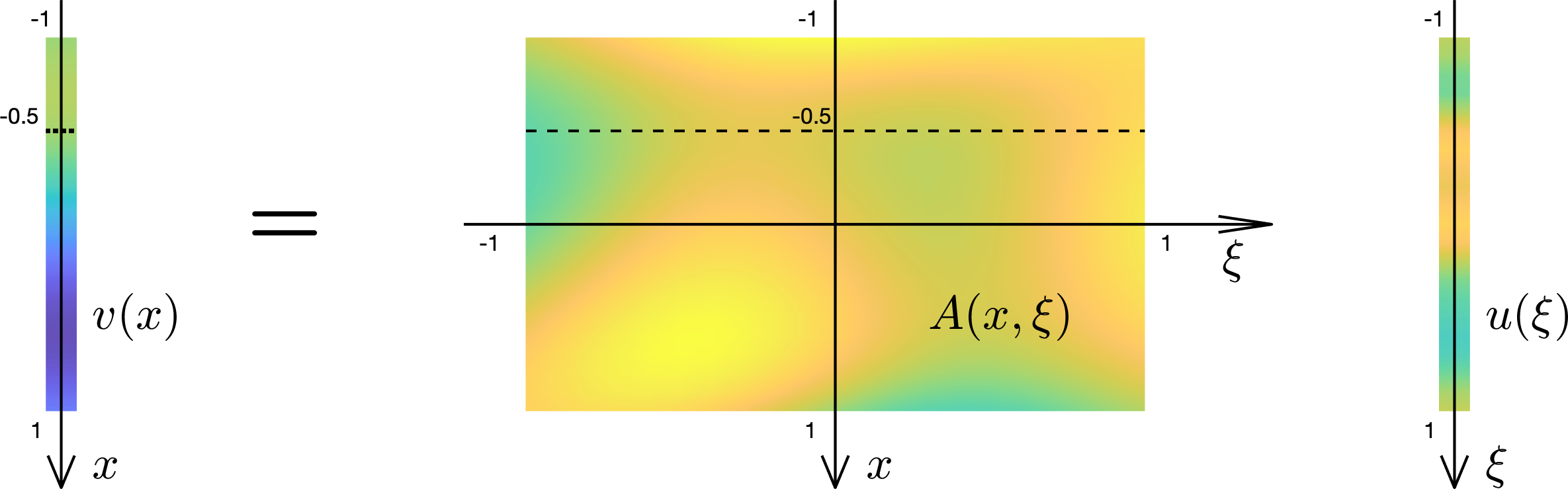} 
            	
            	\mycaption{Graphical depiction of the integral operator~\req{a_kernel} as an abstraction of matrix-vector
            		multiplication. The two-variable kernel function $A(x,\xi)$ is the counterpart of matrix entries, with the coordinate 
            		$x$ as ``row index'' and $\xi$ as ``column index''. The operation $v(x) = \int_{\sm1}^1 A(x,\xi) ~u(\xi)~ d\xi$ gives the value of 
            		$v(x)$ at any $x$ (an instance above is depicted by the dashed lines at $x=-0.5$) as the multiply-then-integrate 
            		of the ``$x$'th row'' of $A(.,.)$ with the all the values of $u(\xi)$ viewed as a ``column vector''. 
            		The case shown above is for an 
            		integral operator on functions defined over the interval $[-1,1]$. 
            		The unusual choice of the
            		 vertical axis positive direction as downwards is made  to be in analogy with matrix rows being
            		indexed from top to bottom.  }
            	\label{kernel_vis.fig}
            \end{figure}
            The operation~\req{a_kernel} is depicted in Figure~\ref{kernel_vis.fig}.
            The one-variable functions $u(\xi)$ and $v(x)$ are analogous to ``column vectors'', while the two-variable
             kernel function $A(x,\xi)$ is analogous to a matrix, i.e.  a two-dimensional array. For each $x$, the value of $v(x)$ is 
             given by the 
             operation of multiply-then-integrate of the corresponding ``row'' of $A(x,\xi)$ with the function $u(\xi)$ in an analogous 
             manner to matrix-vector multiplication.

             Given two operators $A$ and $B$ in terms of their respective kernel functions,  it is easy to see that the operator 
            sum $C:=A+B$ has as its kernel function $C(x,\xi)=A(x,\xi)+B(x,\xi)$ 
            \begin{align*}
            	v &=~ \big( A+B \big) u ~=~ Au ~+~ Bu  		\\ 
            	v(x) &=~ \int A(x,\xi)~u(\xi)~d\xi ~+~  \int B(x,\xi)~u(\xi)~d\xi ~=~ \int \big( A(x,\xi)+B(x,\xi) \big) ~u(\xi)~d\xi. 
            \end{align*} 
            Therefore, under addition,  kernel functions behave just like matrix-matrix addition which is element-by-element. 
            
            Another intuitive  property of  kernel representations is that they can be composed in a manner similar to 
            matrix-matrix multiplication. Let $A:u\mapsto v$ and $B:v\mapsto w$ be two operators with kernel representations 
            \[
            	v(x) ~=~ \int A(x,\xi) ~u(\xi) ~d\xi, 
            	~~~~~~~~~~~~
            	w(x) ~=~ \int B(x,\xi) ~v(\xi) ~d\xi. 
            \]
            Define a third operator as the composition $C:=BA:u\mapsto w$, and calculate its kernel representation from those 
            of $A$ and $B$ as follows 
            \begin{align*}
            	w(x) =  \int B(x,\xi) ~v(\xi) ~d\xi &=   \int B(x,\xi) \lb \int A(\xi,r) ~u(r) ~dr \rb ~d\xi    		\\
            			&=   \int \lb \int B(x,\xi)  ~A(\xi,r) ~d\xi  \rb ~u(r) ~dr     					
            			=   \int C(x,r)  ~u(r) ~dr     .
            \end{align*}
            Thus the kernel of the composition $C=BA$ is obtained from the formula 
            \begin{figure}[t]
            	\centering
            		\includegraphics[width=0.95\textwidth]{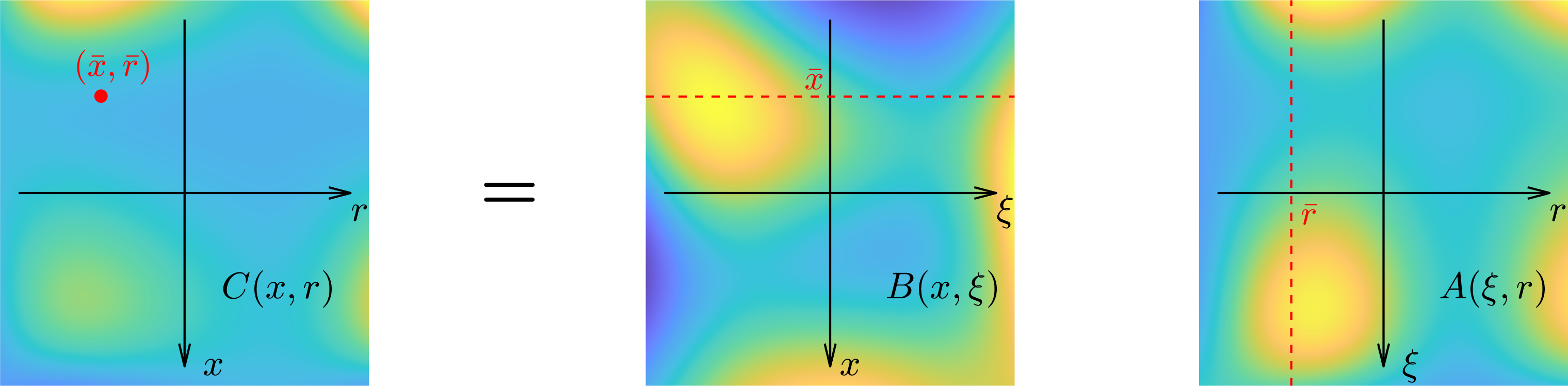} 
            	
            	\mycaption{A graphical depiction of the composition of two operators $C=BA$
            		 as the integral operation~\req{comp_formula} on their respective kernels. This operation
            		 is akin to matrix-matrix multiplication as shown above. The value of the kernel $C$ at a point 
            		 $(\bar{x},\bar{r})$ is obtained from integrating the ``row'' $B(\bar{x},.)$ against the ``column'' 
            		 $A(.,\bar{r})$.   }
            	\label{kernel_composition.fig}
            \end{figure}
            \be
            	C(x,r) ~=~ \int  B(x,\xi)  ~A(\xi,r) ~d\xi ,
              \label{comp_formula.eq}
            \ee
            which looks like matrix-matrix multiplication except for integration instead of summation. 
             Each  ``row'' $B(x,.)$ of the kernel of $B$ is integrated against each 
            ``column'' $A(.,r)$ of the kernel of $A$. 
            The composition operation~\req{comp_formula} 
            is  depicted graphically in Figure~\ref{kernel_composition.fig}. The reader should compare this visually
            with the usual matrix-matrix multiplication.

	 \subsubsection*{Lower-Triangular Operators}

\begin{figure}[t]
            		\centering
			\begin{subfigure}{0.45\textwidth} 
				\centering 
            			\includegraphics[height=.15\textheight]{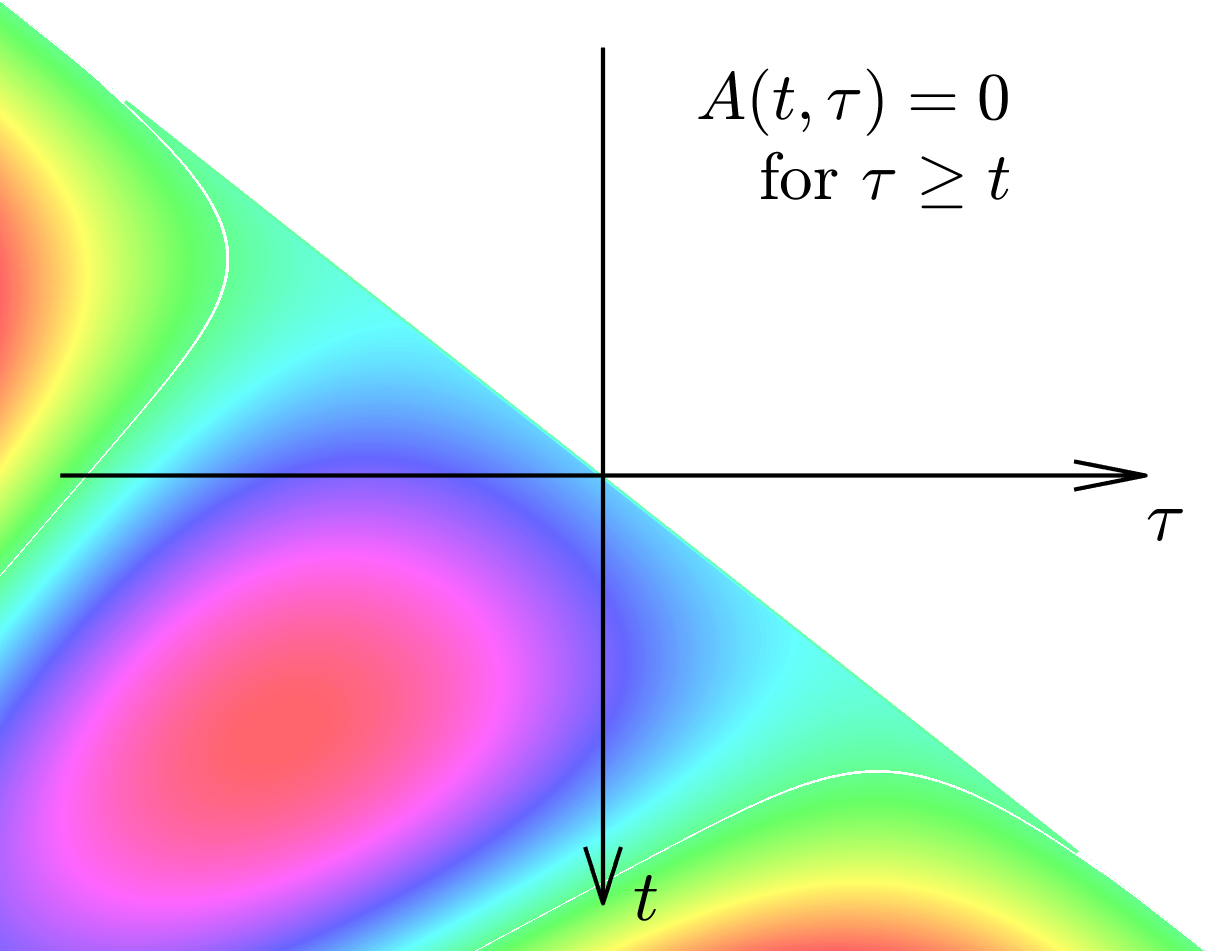} 
				
				\mysubcaption{A {\em lower-triangular kernel} is such that $A(t,\tau)=0$ for $\tau\geq t$. If it operates 
            		   		on time signals, a lower-triangular kernel is a {\em causal system}, i.e. past values of the
            			 	output do not depend on future values of the input.
					} 
            		    \label{kernels_vis_b.fig}
		   	\end{subfigure} 
			\quad\quad
			\begin{subfigure}{0.45\textwidth} 
				\centering 
            			\includegraphics[height=.15\textheight]{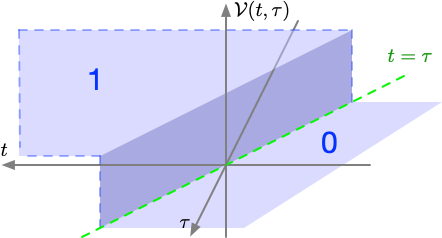} 
				
				\mysubcaption{The kernel function of the  Volterra (forward)  integration operator $\Volt$
					has value one over the lower-triangular region $\tau<t$ and zero everywhere else. 
					It is analogous to a strictly lower-triangular matrix with ones on all subdiagonals. 
					} 
            		    \label{Volterra_vis.fig}
		   	\end{subfigure}

             		   \mycaption{ Graphical depiction of lower-triangular operators and the Volterra forward integration operator. 
		   			} 
            \end{figure}

		Just like certain matrix structures encode  symmetries or properties of the linear operations they represent, 
             the structure of a kernel encodes  properties of the operators they represent. Figure~\ref{kernels_vis_b.fig} 
            illustrates the structure of kernel functions of  what can be termed  ``lower-triangular'' operators. 
		Such operators arise when modeling time-varying {\em causal} systems. 
		The  kernel is restricted to be zero in the ``upper triangular 
		part'' of the $(\tau,t)$ plane 
		\be
			A(t,\tau) ~=~ 0 , ~~~\mbox{for}~ \tau\geq t . 
		  \label{LT_cond.eq}
		\ee	
		If $u$ and $y$ are temporal signals over the entire real line, 
		then the lower-triangular property of the kernel implies that the  integral~\req{a_kernel} has the
		following limits  
		\be
			y(t) ~=~ \int_{-\infty}^\infty A(t,\tau) ~u(\tau)~ d\tau ~=~  \int_{-\infty}^t A(t,\tau) ~u(\tau)~ d\tau.
		  \label{A_low_triang_causal.eq}
		\ee	
		When $t$ and $\tau$ are interpreted as time, then~\req{A_low_triang_causal} is  
		 the description of a general time-varying system mapping $u$ to $y$ that has the causality 
		property, i.e. for any given time $t$, current and past values of the output $\lcb y(\tau); ~\tau\leq t \rcb$ do not depend 
		on future values of the input $\lcb u(\tau); ~\tau> t \rcb$.

		An alternative way of imposing the lower-triangular condition~\req{LT_cond} is by using the 
		{\em unit-step  (Heaviside)} function $\heavi$ as follows. Given any kernel function $A(x,\xi)$, observe that
		the product 
		$A(x,\xi) \step{x-\xi}$ becomes a lower triangular kernel 
		\[
			\int^{\overline{\xi}}_{\underline{\xi}} \big( A(x,\xi) \step{x-\xi} \big) ~u(\xi)~ d\xi 
				~=~  \int_{\underline{\xi}}^x A(x,\xi) ~u(\xi)~ d\xi , 
		\]
		since $\step{x-\xi}=0$ when $\xi>x$. The above holds regardless of the original upper and lower integration limits 
		$\overline{\xi}$ and $\underline{\xi}$ respectively.

		Operators with a lower triangular kernel are sometimes called {\em Volterra operators} if the kernel function 
		is bounded. For Volterra operators acting on function spaces $L^p(\cI)$ where $\cI$ is compact, 
		these operators have the important property that the Neumann series converges even if the operator norm 
		is greater than one. 
		We first investigate a particular Volterra operator, which is the forward-integration operator defined in~\req{int_op_def}.

	\subsection{The Volterra Integration Operator $\Volt$} 
	
	The integration operator~\req{int_op_def}  has a 
	kernel representation in terms of the unit step function as follows 
	\be
		\big( \Volt g \big)(t) ~:=~ \smint{0}{t} g(\tau) ~d\tau ~=~ \smint{0}{T} \heavi(t\sm\tau) ~g(\tau) ~d\tau 
		\hstm \Leftrightarrow \hstm 
		\Volt(t,\tau) ~=~ \heavi(t\sm\tau) , 
	  \label{Volterra_def.eq}
	\ee
	were we used the notation $\Volt(t,\tau)$ for the kernel function of the operator $\Volt$. 
	This operator is analogous to a
	strictly  lower-triangular matrix where all the entries below the diagonal are  
	$1$. This is illustrated in Figure~\ref{Volterra_vis.fig}. 
	
	A strictly lower-triangular matrix is  nilpotent, i.e. the first $k$ subdiagonals of the $k+1$ power of the 
	matrix is zero, and thus it becomes zero after raising to a sufficiently large power. 
	Although the operator $\Volt$ is not nilpotent, it 
	 does becomes 
	``smaller''  as it is composed with itself repeatedly, so it can be thought of as {\em asymptotically nilpotent}. 
	 More precisely, the composition formula~\req{comp_formula}  for operator 
	kernels implies that 
	\[
		\Volt^2(t,\tau) ~=~ 
		\smint{0}{T} \Volt(t,r) ~\Volt(r,\tau) ~ dr ~=~ 
		\smint{0}{T} \heavi(t\sm r) ~\heavi(r\sm\tau) ~ dr 
		~=~ \smint{\tau}{t} dr ~=~ (t-\tau) ~\heavi(t\sm\tau) .
	\]
	Repeated applications of this calculation~\cite{riesz2012functional} show that\footnote{As can be verified by induction.}  
	\be
		\Volt^k(t,\tau) ~=~ \frac{(t-\tau)^{k-1}}{(k-1)!} ~ \heavi(t\sm\tau) . 
	 \label{Volterra_k.eq}
	\ee
	Note that for each $(t,\tau)$, the  kernel of $\Volt^k$ limits to zero as $k\rightarrow\infty$ 
	 since the factorial in the denominator grows faster than any 
	power of $k$. This is the operator counterpart of a strictly lower triangular matrix being nilpotent, and we call 
	this property {\em asymptotic nilpotence}. 
	Asymptotic nilpotence implies 
	 that the Neumann series expression converges in the operator norm (on $\LP{p}[0,T]$) ($p\in[1,\infty]$) 
	  as outlined in 
	Appendix~\ref{Neu_conv.app}. 
	
	The expression~\req{Volterra_k} gives a useful formula 
	for repeated integration of any function. Define the  {\em $k$'th antiderivative} of any function $g$ by 
	\[
		g^{(\sm k)} (t) ~:=~ \smint{0}{t} \smint{0}{\tau_{k}} \cdots \smint{0}{\tau_2} g(\tau_1) ~d{\tau_{1}} 
					~\cdots~ d{\tau_{k\sm1}} ~d{\tau_k} 
		\hstm \Leftrightarrow \hstm 
		g^{(\sm k)} ~:=~ \Volt^k g, 
	\]
	and note the consistency of this notation with that for the $k$'th derivative of a function. Applying the expression~\req{Volterra_k}
	for the kernel of $\Volt^k$ we see that 
	\be
		g^{(\sm k)} (t) 
		~=~ 
			\frac{1}{(k-1)!} 
		\int_{0}^{t}  (t-\tau)^{k-1}  ~ g(\tau) ~d\tau 
	  \label{Cauchy_rep_int.eq}
	\ee
	This formula is known as the {\em Cauchy formula for repeated integration}. 
	One interesting application of this formula is to define
	``fractional integrals'' where the  non-negative integer $k$  is replaced by a non-negative real number. The term 
	$(t-\tau)^{(k-1)}$ would still make sense, and the term $(k-1)!$ is replaced by the Gamma function $\Gamma(k-1)$. 
	We will not need the concept of fractional integration in this note.

	\section{Formulas for the State Transition Matrix} 						\label{Form_STM.sec}
	
	In this section, we consider linear time-varying systems without input of the form 
	\be
		\xd(t) ~=~ A(t)~ x(t), 
		\hstm \hstm t\in[0,T] ,
	  \label{LTV_STM_form.eq}
	\ee
	and calculate the Neumann series expression for the response due to initial conditions only. 
	For notational simplicity, we assume temporarily that the initial condition is given at $t=0$, 
	and write $\Phi(t)$ for $\Phi(t,0)$.  
	By Lemma~\ref{STM.lemma} the solution to~\req{LTV_STM_form} is given 
	 in terms of the state transition matrix, which is the 
	solution to the matrix  differential equation
	\[
		\dot{\Phi}(t) ~=~ A(t) ~\Phi(t) , 
		\hstm 
		\Phi(0) = I, 
		\hstm\hstm 
		t\in[0,T]. 
	\]
	Just like the vector case, this equation can be written as an integral equation 
	\[
				\Phi ~=~ \Volt \cA ~\Phi ~+~ \heavi I 
				\hstm \Leftrightarrow \hstm 
				\big( I - \Volt\cA \big) ~\Phi ~=~ \heavi I , 
	\]
	where $I$ is the identity matrix, and the function $\big(\heavi I \big) (t) = I,$ for $t\in[0,T]$. 
	The abstract formula for 
	the solution again follows from the Neumann series
	\be
		\Phi ~=~ \big( I  - \Volt \cA \big)^{-1}   \heavi I 
		~=~\big(    I ~+~ \Volt \cA ~+~ \big( \Volt \cA \big)^2 ~+~ \cdots  \big) ~\heavi I
	  \label{Neumann_LTV_hI.eq}
	\ee
	Our goal is to express this series in terms of the system parameter $A(.)$. 
	This is done for the time-invariant case first, where the Neumann series will yield the exponential 
	function,  and then generalized to the time-varying case, which will yield the less explicit 
	Peano-Baker series.

	\subsection{The Time-invariant Case: The Exponential Function} 
	
	In the time-invariant case,  $A(t)$ is constant in $t$ (so we just denote it by $A$). 
	The action of  the operator $\Volt\cA$ on any 
	function $g$ is 
	\[
		\big( \Volt\cA ~g\big) (t) 
		~=~ 
		\int_0^t A~ g(\tau) ~d\tau 
		~=~ A \int_0^t g(\tau) ~d\tau 
		~=~ \big( \cA \Volt ~g \big) (t) . 
	\]	
	Thus in the time-invariant case, 
	 the operators $\Volt$ and $\cA$ commute  ($\Volt\cA=\cA\Volt$), and this makes the calculation of 
	the Neumann series~\req{Neumann_LTV_hI}  particularly easy
	\begin{align} 
		 \big( I  - \Volt \cA \big)^{-1} 
		 ~&=~ I ~+~ \Volt \cA ~+~  \Volt \cA  \Volt \cA  ~+~  \Volt \cA  \Volt \cA  \Volt \cA  ~+~ \cdots  			\nonumber 	\\ 
		 ~&=~ I ~+~ \cA\Volt  ~+~  \cA^2 \Volt^2   ~+~  \cA^3  \Volt^3   ~+~ \cdots
		 ~=~ \sum_{k=0}^\infty \cA^k \Volt^k.
	  \label{Neumann_comm_abs.eq}
	\end{align}
	Now  compute the kernel representation of the operator $\big( I - \Volt\cA \big)^{-1}$.
	Using~\req{Volterra_k}, and noting that the operator 
	$\cA^k$ is simply multiplication by the matrix $A^k$, we see that 
	\be
		 \big( I  - \Volt \cA \big)^{-1}(t,\tau) ~=~ \sum_{k=0}^\infty A^k \Volt^k(t,\tau) 
		 ~=~ 
		 \sum_{k=0}^\infty A^k \frac{(t-\tau)^{k-1}}{(k-1)!} ~ \heavi(t\sm\tau) 
	 \label{Neumann_commute.eq}
	\ee
	Applying this to the solution formula~\req{Neumann_LTV_hI}
	\begin{align*} 
		\Phi(t) ~&=~ 
		\lb \big( I  - \Volt \cA \big)^{\sm1}  ~\heavi I \rb(t) 				
		~= ~
		\int_0^T  \big( I  - \Volt \cA \big)^{-1}(t,\tau) ~\big( \heavi I \big)(\tau) ~d\tau 						\\
		~&= ~
		 \sum_{k=0}^\infty A^k \int_0^T \frac{(t-\tau)^{k-1}}{(k-1)!} ~ \heavi(t\sm\tau) ~ \heavi(\tau) I ~d\tau 		\\
		~&= ~
		 \sum_{k=0}^\infty A^k \int_0^t \frac{(t-\tau)^{k-1}}{(k-1)!}   ~d\tau 						
		 ~=~ 
		 \sum_{k=0}^\infty A^k~ \left. \frac{\sm(t-\tau)^{k}}{k!} \right|_{0}^{t} 						
		 ~=~ 
		  \sum_{k=0}^\infty \frac{ A^k  t^k}{k!} 									
		  ~=:~ e^{At}  . 
	\end{align*} 
	This is exactly the solution as postulated in~\req{exp_At_sol} earlier. However, in this case, the matrix exponential 
	$e^{At}$ emerges naturally (without guessing) from the details of the Neumann series for the time-invariant setting.

	\subsection{Time-varying Systems: The Peano-Baker Series}

	For this calculation and for the subsequent one with non-zero input, it will be useful to 
	switch notation, and 
	derive the expressions for the state transition matrix for a general initial time $\tau$
	\be
		\dot{\Phi}(t,\tau) ~=~ A(t) ~\Phi(t,\tau), 
		\hstm
		\Phi(\tau,\tau)=I, 
		\hstm\hstm 
		0\leq \tau\leq t \leq T. 
	  \label{LTV_STM_PB_tb.eq}
	\ee
	In this setting, the Volterra integration operator is the forward integration operator starting at time $\tau$ 
	\[
		\big( \Volt ~g \big) (t) ~:=~ \smint{\tau}{t} g(\tau_1) ~d\tau_1. 
	\]
	In the general time-varying case,   the operators $\cA$ and $\Volt$
	are  no longer necessarily commutative, and the Neumann series cannot be rearranged into
	 the simpler form~\req{Neumann_comm_abs}. 
	For notational simplicity,  relabel the composition 
	\begin{align}
		\hspace{-1.5em} 
		\Volt_A := 
		\Volt \cA  
		\hstm\Leftrightarrow \hstm 
		\big( \Volt_A ~g \big) (t) &:= \smint{\tau}{t} A(\tau) ~g(\tau_1)~ d\tau_1 
			= \smint{0}{T} A(\tau_1)\heavi(t\sm\tau_1)\heavi(\tau_1\sm\tau)   ~g(\tau_1)~ d\tau_1 
																\nonumber			\\ 
		\Rightarrow \hstm 
		\Volt_A(t,\tau_1) ~&=~ A(\tau_1) \heavi(t\sm \tau_1)\heavi(\tau_1\sm\tau)  ,
																\label{V_A_first.eq}
	\end{align} 
	where the last expression is for the kernel function of the operator $\Volt_A$. Note that the independent variables 
	in this kernel function are $(t,\tau_1)$, while $\tau$ should be regarded as a parameter specifying the initial condition 
	time (and therefore a fixed number when applying the operator $\Volt_A$). 
	This notational switching will turn out to significantly simplify subsequent notation. 
	
	Expressions for the kernels of powers $\Volt_A^k$ can get notationally messy as they will involve multivariable 
	integrals. The notation will be significantly simplified by the introduction of the {\em multivariable Heaviside 
	 function}
	 \[
		 \heavi_{t_1,t_2,\ldots,t_n} 
		 ~:=~ 
		 \left\{ \begin{array}{ll}
		 	1, & t_1\geq t_2 \geq \cdots \geq t_n 		\\ 
			0, & \mbox{otherwise.} 
		\end{array} 	\right. .
	 \]
	 This is just  convenient and compact notation for the product of several scalar Heaviside functions, 
	 which can be  used as an alternative  definition 
	 \[
	 	 \heavi_{t_1,t_2,\ldots,t_n} 
		 ~:=~ 
		 \heavi(t_1-t_2) ~\heavi(t_2-t_3)~ \cdots ~ \heavi(t_{n\sm1} - t_n ) . 
	 \]
	 This function allows for encoding integration limits in the following manner 
	 \begin{multline} 
	 	\smint{0}{T} \cdots \smint{0}{T} f(t,\tau_1,\ldots, \tau_n,\tau) ~
	 		\heavi_{t,\tau_1,\ldots,\tau_n,\tau} ~d\tau_1 \cdots d\tau_n  			\\
		~=~ 
	 	\smint{\tau}{t}    \smint{\tau}{\tau_1}     \cdots   \smint{\tau}{\tau_{n\sm2}}    \smint{\tau}{\tau_{n\sm1}} 
			f(t,\tau_1,\ldots, \tau_n,\tau) ~
	 		 ~d\tau_n \cdots d\tau_1 			
			\hstm 
			t,\tau\in[\tb,T]
	  \label{heavi_int_prop.eq}
	 \end{multline} 
	 Finally observe that the multivariable Heaviside function obeys the following ``concatenation'' 
	 property which will simplify later manipulations 
	 \be
	 	\heavi_{t_1,\ldots,t_{k\sm1},t_k,t_l} ~\heavi_{t_k,t_{k+1},\ldots,t_n} 
				~=~  \heavi_{t_1,\ldots,t_{k\sm1},t_k,t_{k+1}, \ldots,t_n} , 
		\hstm 
		\mbox{if} ~ l=k+1,\ldots, n. 
	  \label{m_heavi_conc.eq}
	 \ee

	We now return to the calculation of the kernel functions of the operators $\Volt_A^k$. In the new notation, 
	the kernel calculated in~\req{V_A_first} becomes 
	$\Volt_A(t,\tau_1)~=~ A(\tau_1) \heavi_{t,\tau_1,\tau}$, where $\tau$ is a fixed number denoting the initial 
	condition time. 
	Calculations of subsequent powers give 
	\begin{align*} 
		\hspace{-2em} 
		\big( \Volt_A^2\big) (t,\tau_1) 
		&= 
                		\smint{0}{T} \Volt_A(t,\tau_2) ~\Volt_A(\tau_2,\tau_1) ~d\tau_2 
                		= 
                		\smint{0}{T} A(\tau_2) \heavi_{t,\tau_2,\tau} ~A(\tau_1) \heavi_{\tau_2,\tau_1,\tau}  ~d\tau_2 		\\
                		&= 
                		\smint{0}{T} A(\tau_2)~  \heavi_{t,\tau_2,\tau_1,\tau}  ~d\tau_2 	~~A(\tau_1), 			\\
		\hspace{-2em} 
		\big( \Volt_A^3\big) (t,\tau_1) 
		&= 
            		\smint{0}{T}  \! \Volt_A(t,\tau_3) ~ \Volt_A^2(\tau_3,\tau_1) ~ d\tau_3 		\\
            		&= 
            		\smint{0}{T} \!
            			A(\tau_3) \heavi_{t,\tau_3,\tau}	
				\lb  \smint{0}{T} A(\tau_2)~  \heavi_{\tau_3,\tau_2,\tau_1,\tau}  ~d\tau_2 ~A(\tau_1) \rb	d\tau_3																										\\
            		&= 
            		\smint{0}{T} \smint{0}{T} 
            			A(\tau_3)A(\tau_2)~  \heavi_{t,\tau_3,\tau_2,\tau_1,\tau}  ~d\tau_2 d\tau_3	~~A(\tau_1), 
	\end{align*} 
	Note the use of
	 the concatenation  property~\req{m_heavi_conc} to simplify the final expression. 
	Repeated applications of this calculation show that 
	\be
		\big( \Volt_A^k \big)(t,\tau_1)  
		~=~ 
		\smint{0}{T}\cdots \smint{0}{T} 
			 A(\tau_k)\cdots A(\tau_{2})~  \heavi_{t,\tau_k,\ldots,\tau_{2},\tau_1,\tau}  
			 	~d\tau_2 \cdots d\tau_{k}  	~~A(\tau_1) . 
	  \label{PB_k_term.eq}
	\ee

	Now we turn to the evaluation of the series~\req{Neumann_LTV_hI} for the state transition matrix. The 
	$k$'th element of that series, which we denote by $\Phi_k$ is calculated by 
	\begin{align} 
		\Phi_k(t,\tau) ~&:=~ \big(  \Volt_A^k~ \heavi I \big)(t)  
			~=~ \smint{0}{t} \Volt_A^k(t,\tau_1) ~\heavi_{\tau_1,\tau} ~I ~d\tau_1  
			~=~ \smint{\tau}{t} \Volt_A^k(t,\tau_1) ~d\tau_1  			\nonumber		\\
			&=~ 
		\smint{\tau}{t} 
		\smint{0}{T}\cdots \smint{0}{T} 
			 A(\tau_k)\cdots A(\tau_{2})~  \heavi_{t,\tau_k,\ldots,\tau_2,\tau_1,\tau}  
			 	~d\tau_2 \cdots d\tau_{k}  	~A(\tau_1)  ~d\tau_1 				\nonumber		\\
			&{=}~ 
		\smint{0}{T}\cdots \smint{0}{T} 
			 A(\tau_k)\cdots A(\tau_1) ~  \heavi_{t,\tau_k,\ldots,\tau_1,\tau } 
			 	~d\tau_2 \cdots d\tau_{k}  ~d\tau_1								\nonumber		\\
			&{=}~ 
		\smint{\tau}{t}
			 A(\tau_k) \smint{\tau}{\tau_k} \cdots \smint{\tau}{\tau_{2}}  A(\tau_1)
			 	~ d\tau_{1} \cdots d\tau_k, 								\label{Phi_k_expression.eq}
	\end{align} 
	where the propery ~\req{heavi_int_prop}
	of $\heavi$ gives the integration limits in the last equation. 
	
	We finally conclude that the state transition matrix $\Phi(t,\tau)$
	of the system~\req{LTV_STM_form} 
	is given by the {\em Peano-Baker  series} 
	\be
		\Phi(t,\tau) ~=~ I ~+~ \Phi_1(t,\tau) ~+~ \Phi_2(t,\tau) ~+~ \cdots , 
	  \label{PB_series.eq}
	\ee
	where each $\Phi_k$ is given by~\req{Phi_k_expression}. 
	Note that the expression~\req{Phi_k_expression} also implies that the series terms 
	have the following recursion relationship
	\[
		\Phi_k(t,\tau) = \smint{\tau}{t} A(\tau_k) ~\Phi_{k\sm1}(\tau_k,\tau) ~d\tau_k 
		\hsom \Leftrightarrow \hsom 
		\dot{\Phi}_k(t,\tau) = A(t) ~\Phi_{k\sm1}(t,\tau), 
		\hsom 
		\Phi_k(\tau,\tau) = 0, ~ k\geq 1,   
	\]
	with $\Phi_0(t,\tau) =I$. 
	The convergence of the Peano-Baker series is a consequence of the ``asymptotic nilpotence'' of the 
	Volterra integration operator. Appendix~\ref{Neu_conv.app} details this argument. 
	
	In general, the Peano-Baker series terms~\req{Phi_k_expression} 
	do not yield tractable expressions except in special cases. 
	One such case is when the time-varying family of matrices $\lcb A(t), ~t\in[0,T] \rcb$ mutually commute. In this case, 
	the series can be used to express $\Phi$ in terms of a matrix exponential. 
	\begin{lemma}														\label{PBS_commutative.lemma} 
		Consider a mutually commuting family of matrices $\lcb A(t), ~t\in[0,T] \rcb$. The Pean-Baker series~\req{PB_series}
		for the state transition matrix reduces to 
		\be
			\Phi(t,\tb) ~=~ \exp\lb \int_\tb^t A(\tau) ~d\tau \rb . 
		  \label{STM_commute.eq}
		\ee	
	\end{lemma}
	The proof of this lemma is in Appendix~\ref{PBS_commutative.appen}. The basic idea is that commutativity allows for 
	expressing the repeated integral/products in~\req{PB_k_term} as a power of a single integral. The Peano-Baker 
	series then becomes 
	the series for the matrix exponential of the integral above. 
	We finally note that in the scalar case, the formula~\req{STM_commute} can be derived directly by integrating 
	the equations as shown in Exercise~\ref{STM_scalar.ex}.

	\section{Systems with Inputs} 											\label{inputs.sec}
	
	We will show that the solution to the linear time-varying system with input
	\be
		\xd(t) ~=~ A(t) ~x(t) ~+~ w(t), 
	  \label{LTV_VOC_form.eq}
	\ee
	 is given by the so-called 
	{\em variations of constants} formula 
	\be
		x(t) ~=~ 
			\underbrace{~\Phi(t,\tb) ~x(\tb)~}_{\mbox{\footnotesize zero-input response}}
			~+~
				\underbrace{~\int_{\tb}^t \Phi(t,\tau) ~w(\tau) ~d\tau,~}_{\mbox{\footnotesize input-to-state response}}
	 \label{VOC_LTV.eq} 
	\ee
	where $x(\tb)$ is an initial condition, and $\Phi$ is the state transition matrix of the homogenous 
	problem (i.e. the problem with $w(t)=0$).  This formula is states that the solution is the sum 
	of $\Phi(t,\tb) ~x(\tb)$, which is the {\em response due to the initial condition} (also called the 
		{\em zero-input response}), and the {\em 
	response due to input} (also called the {\em input-to-state response}), 
	which is a linear operation on $\lcb u(\tau), ~\tau\in[\tb,t] \rcb$, 
	the input function restricted to  the time interval $[\tb,t]$.

	The formula~\req{VOC_LTV} can be directly verified by differentiation. 
	First recall the Leibniz integral rule, which  is the fundamental theorem of calculus when the integral limits depend
	on the differentiation variable (see Exercise~\ref{Leibniz.ex} for a proof). 
	In this specific case it states that for any function $f$ of two variables
	\be
			\frac{d}{dt} \int_{\tb}^{t} f(t,\tau) ~d\tau ~=~ \int_\tb^t \frac{\partial}{\partial t} f(t,\tau) ~d\tau
					~+~ f(t,t) .
	  \label{Leibniz_rule.eq}
	\ee
		If $f$ is matrix-valued, this formula applies entry by entry.	Now differentiating~\req{VOC_LTV}
	\begin{align} 
		\hspace{-2em}
		\xd(t) ~&=~  \frac{d}{dt} \lb \Phi(t,\tb) ~x(\tb) + \smint{\tb}{t} \Phi(t,\tau) ~w(\tau) ~d\tau \rb	
																	\nonumber 		\\
			~&=~  A(t) ~{\Phi}(t,\tb)   ~x(\tb) +  \smint{\tb}{t} \frac{\partial}{\partial t} \Phi(t,\tau) ~w(\tau) ~d\tau 	
									+ \Phi(t,t) ~w(t) 
																	\nonumber 		\\
			~&=~  A(t)~ \lb \Phi(t,\tb) ~x(\tb)  +    \smint{\tb}{t} \Phi(t,\tau) ~w(\tau) ~d\tau \rb
									~+~   w(t) 
				~=~ A(t) ~x(t) ~+~ w(t). 									\label{VOC_verif.eq}
	\end{align} 

	While the formula~\req{VOC_LTV} is relatively easy to verify, it is not clear where it comes from or how 
	one can discover it from first principles. In the following we present three different methods of arriving at 
	this formula from basic principles. Each method gives additional insight into the problem. First, we consider 
	the response of the system for a special input which is a Dirac delta function. 
	
\begin{figure} 
		\centering
            	\begin{subfigure}[b]{0.45\textwidth}  
            		\centering 
            		\includegraphics[width=0.9\textwidth]{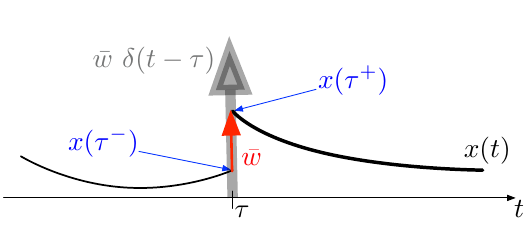} 
            		
            	  \mysubcaption{The input $\wba ~\delta(t-\tau)$ is an impulse at time 
            	  	$\tau$ with  	strength given by the vector $\wba$. This input causes the state to ``jump'' 
            		from $x(\tau^-) := \lim_{t\nearrow\tau}$ to  $x(\tau^+) := \lim_{t\searrow\tau}$. The magnitude and 
            		direction of the jump is given by the vector $\wba$.  } 
            	 \label{state_discont_delta.fig}
            	\end{subfigure} 	
		\quad
            	\begin{subfigure}[b]{0.45\textwidth}  
            		\centering 
            		\includegraphics[width=0.9\textwidth]{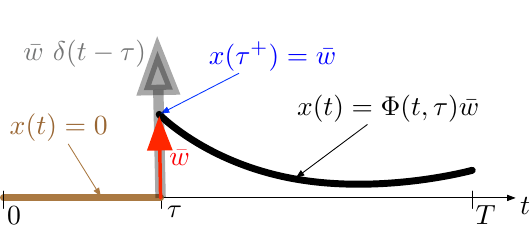} 
            		
            	  \mysubcaption{When the input is an impulse $\wba ~\delta(t-\tau)$  at time $
	  			\tau\in(0,T)$, with zero initial 
	  			conditions $x(0)=0$, the state becomes non-zero at $\tau$, and evolves
				as if $x(\tau)=\wba$ is an initial condition. The entire solution is then 
				$x(t) ~=~\Phi(t,\tau)~\wba~\heavi(t-\tau)$. 
	  			} 
            	 \label{state_impulse_in.fig}
            	\end{subfigure} 	
	
		\mycaption{Illustration of the behavior of a linear system~\req{state_in_delta} when 
            	  	the input is an impulse.
			} 
	  \label{LTV_impulse.fig} 
	\end{figure}

	Consider an input of the form $w(t) =\wba ~\delta(t-\tau)$, a delta function in the direction of the vector $\wba$
	  applied at time $\tau$
	\be
		\xd(t) ~=~ A(t) ~x(t) ~+~ \wba~\delta(t-\tau) . 
	  \label{state_in_delta.eq}
	\ee
	To see what happens around the time $\tau$, integrate the equation over $[\tau-\epsilon,\tau+\epsilon]$ 
	\begin{align*} 
		~
		\smint{\tau-\epsilon}{\tau-\epsilon}	\xd(t) ~dt
		~&=~ \smint{\tau-\epsilon}{\tau-\epsilon} A(t) ~x(t) ~dt ~+~ \smint{\tau-\epsilon}{\tau-\epsilon} \wba~\delta(t-\tau)~ dt	\\
		x(\tau+\epsilon) - x(\tau-\epsilon) 
		~&=~ \smint{\tau-\epsilon}{\tau-\epsilon} A(t) ~x(t) ~dt ~+~ \wba.
	\end{align*} 
	Provided that the function $A(t)x(t)$ is bounded, the last integral term becomes zero when taking
	the limit $\epsilon\rightarrow 0$, and we conclude that 
	\be
		x(\tau^+) ~=~ x(\tau^-) ~+~ \wba , 
		\hstm \hstm\hstm 
		\arraycolsep=3pt
		\begin{array}{rcl} 
			x(\tau^+) &:= & \lim_{t\searrow\tau} x(t) \\ 
			x(\tau^-) &:= & \lim_{t\nearrow\tau} x(t)
		\end{array} .
	  \label{imp_state_disc.eq}
	\ee
	Thus the effect of a delta function at time $\tau$ in the input is to make the state ``jump'' 
	from $x(\tau^-)$ just before $\tau$ to $x(\tau^+)$ just after $\tau$, with the jump magnitude and direction equal to the 
	vector $\wba$. This is illustrated in Figure~\ref{state_discont_delta.fig}. 
		
	For example, consider  the system~\req{state_in_delta}  starting from zero initial conditions $x(0)=0$, 
	and the impulse is 
	applied at some time $\tau\in(0,T)$. The response is $x(t)=0$ for $t\in[0,\tau)$. Around $t=\tau$ the state  jumps to 
	 $x(\tau^+)=\wba$. Since the input is then zero over the remainder of the time interval $(\tau,T]$, 
	  the state evolves according to $\Phi(t,\tau)~\wba$ since $x(\tau^+)=\wba$ is the initial condition at $t=\tau$. 
	The full evolution over the entire interval $[0,T]$ can then be written as 
	\be
		x(t) ~=~ \Phi(t,\tau)~\wba ~ \heavi(t-\tau), 
		\hstm\hstm 
		t\in[0,T]. 
	  \label{state_tau_evol.eq}
	\ee
	Figure~\ref{state_impulse_in.fig} illustrates this example which we will use next in a superposition argument.

	\subsection{The Variations of Constants Formula  via Superposition} 
	
	The system~\req{LTV_VOC_form} has an input $w$ that is persistently (in time) acting on it. 
	If the signal $w$  can be written as a linear combination of ``simpler'' inputs, for which the 
	solution is already known, then again by linearity we can write the response as a linear
	combination of the individual responses. Consider writing any signal as a integral 
	involving the delta function 
	\be
		w(t)~ =~ \smint{0}{T} \delta(t-\tau) ~w(\tau) ~d\tau ~=:~ \smint{0}{T} \delta_\tau(t) ~w(\tau) ~d\tau. 
	  \label{w_integ_delta.eq}
	\ee
	This integral can be though of as ``weighted sum'' of a parametrized family of delta functions 
	\[
		\lcb \delta_\tau(t):= \delta(t-\tau), ~t\in[0,T]	\rom  \rcb
	\] 
	with the function $w(\tau)$ acting as the 
	``weighting function''. 
	The calculation~\req{state_tau_evol} already gives us the response 
	 to each $\delta_\tau(t) w(\tau)$. We label that response as $x_\tau$ 
	\[
		x_\tau(t) ~=~ \Phi(t,\tau) ~w(\tau) ~\heavi(t-\tau) 
	\]
	Note that this formula should be read so that it is a relation
	between functions of $t$, with $\tau$ as a parameter. 
	
	The response to the ``combined'' signal~\req{w_integ_delta} 
	is then the integral of all of those individual responses 
	\[
		x(t) ~=~\smint{0}{T} x_\tau(t) ~d\tau 
			~=~ \smint{0}{T} \Phi(t,\tau) ~w(\tau) ~\heavi(t-\tau) ~d\tau
			~=~  \smint{0}{t} \Phi(t,\tau) ~w(\tau)  ~d\tau.
	\]
	This is precisely the input-to-state response portion of the variations of constants
	formula~\req{VOC_LTV}.

	\subsection{Linearity of the Input-to-State Response} 
	
	We have already seen in Section~\ref{STM.sec} that the zero-input response is a linear 
	mapping from initial conditions to the response at any time. It is similarly easy to show 
	that with zero initial conditions, the input-to-state response must be linear. 
	Consider two inputs acting on the same system with zero initial conditions 
	 \begin{align*} 
	 	\xd_1(t)  ~&=~ A(t) ~x_1(t)  ~+~ w_1(t) , 
				\hstm \hstm x_1(\tb) ~=~ 0, 			\\
	 	\xd_2(t)  ~&=~ A(t) ~x_2(t)  ~+~ w_2(t) , 
				\hstm \hstm x_2(\tb) ~=~ 0. 
	 \end{align*} 
	Adding both sides of the equations as an arbitrary linear combination shows that
	\begin{multline*} 
		\frac{d}{dt} \big( \alpha x_1(t) + \beta x_2(t) \big) 
		~=~ 
		A(t) ~ \big( \alpha x_1(t) + \beta x_2(t) \big) ~+~ \big( \alpha w_1(t) + \beta w_2(t) \big) 	,	\\
		 	 \alpha x_1(0) + \beta x_2(0)   ~=~ 0. 
	\end{multline*} 
	Thus the response to a linear combination of the two inputs is the same linear combination of 
	their respective responses (when initial conditions are zero). 
	
	Recall the kernel representation of linear operators, by which 
	 any linear mapping of functions on an interval $[0,T]$ to other functions on $[0,T]$ 
	can be written in the form  
	\be
		x(t) ~=~ \smint{0}{T} G(t,\tau) ~w(\tau) ~d\tau  
			 ~=~ \smint{0}{t} G(t,\tau) ~w(\tau) ~d\tau, 
	  \label{ItS_resp_linear.eq}
	\ee
	where the kernel function $G(.,.)$ may contain generalized functions. The first form is general, while 
	the second is for the case when the operator is {\em causal}. This is the case for the 
	system~\req{LTV_VOC_form}
	solved forward in time, as the response $x$ cannot anticipate future values of the input $w$. 
	
	Given that the response is of the form~\req{ItS_resp_linear}, we now can determine what the 
	original differential equation~\req{LTV_VOC_form} implies about the kernel function $G(.,.)$
	\begin{align} 
		\xd(t) ~&=~A(t) ~x(t) ~+~ w(t) 										\nonumber	\\ 
		\frac{d}{dt} \lb  \smint{0}{t} G(t,\tau) ~w(\tau) ~d\tau \rb 
				&=~ A(t)  \lb  \smint{0}{t} G(t,\tau) ~w(\tau) ~d\tau \rb ~+~ w(t)	\nonumber	\\
		  ~\smint{0}{t} \frac{\partial}{\partial t} G(t,\tau) ~w(\tau) ~d\tau  ~+~	G(t,t) ~w(t) 
				&=~ A(t)  \lb  \smint{0}{t} G(t,\tau) ~w(\tau) ~d\tau \rb ~+~ w(t), 	\label{VOC_Leib_one.eq}
	\end{align} 
	where the last equation follows from applying the Leibniz integral rule~\req{Leibniz_rule}. 
	To see what $G(t,t)$ 
	should be, note that the kernel representation implies that 
	\[
		x(t) ~=~ \smint{0}{T} G(t,\tau) ~\wba~\delta(\tau-t) ~d\tau ~=~ G(t,t)~ \wba. 
	\]
	Thus we can determine $G(t,t)$ by applying a delta function input at time $t$ with initial 
	conditions $x(t^-)=0$ and then $G(t,t)=x(t^+)$ is the value of the immediate state response. 
	The formula~\req{imp_state_disc} implies that $x(t^+)= \wba$, and since $\wba=G(t,t)\wba$ 
	for all possible vectors $\wba$, then $G(t,t)$ must be the identity matrix. 
	Applying this to~\req{VOC_Leib_one} we see that 
	\[
		  ~\smint{0}{t} \frac{\partial}{\partial t} G(t,\tau) ~w(\tau) ~d\tau  ~+~ w(t) 
			~=~ A(t)  \lb  \smint{0}{t} G(t,\tau) ~w(\tau) ~d\tau \rb ~+~ w(t). 
	\]
	Since this formula has to hold for all possible input functions $w$, we finally conclude that 
	\[
		\frac{\partial}{\partial t} G(t,\tau) ~=~ A(t) ~ G(t,\tau) , 
		\hstm\hstm 
		G(\tau,\tau) ~=~ I. 
	\]
	This is precisely the differential equation~\req{STM_diffeq} for the state 
	transition matrix found earlier. We therefore  conclude that $G(t,\tau)=\Phi(t,\tau)$, 
	and the linear operation~\req{ItS_resp_linear} can now be rewritten as 
	\[
		x(t) ~=~  \int_0^t G(t,\tau) ~w(\tau) ~d\tau ~=~  \int_0^t \Phi(t,\tau) ~w(\tau) ~d\tau . 
	\]
	Again, this is the input-to-state portion of the  variations of constants
	formula~\req{VOC_LTV}.

	\subsection{The Variations of Constants Formula  via the Neumann Series} 
	
	In calculating the initial-condition response for 
	 general time-varying system,  we used the Neumann series to arrive at the Peano-Baker 
	 series. More precisely, we used the kernel representation~\req{PB_k_term} for each term $\Volt_A^k$
	  in the 
	 Neumann series, and then applied it to constant functions  to give each term~\req{Phi_k_expression} 
	 of the  Peano-Baker series. For the input-to-state response, we return to the 
	 kernel representation~\req{PB_k_term} of $\Volt_A^k$, but apply it to non-constant functions of the form $\Volt ~w$. 

	Recall the abstract formula~\req{SS_sol_Volt}  for the solution, and 
	consider only the input-to-state response (i.e. $\xb=0$), 
	\[
		x ~=~ \big( I  - \Volt_A \big)^{-1}  ~ \Volt ~w 
		~=~  
		\big(    I ~+~ \Volt_A ~+~  \Volt_A^2 ~+~ \cdots  \big)~ \Volt ~w
	\]
	(recall that $\Volt_A:=\Volt \cA$). 
	Each term in this series can be calculated using the kernel function of $\Volt_A^k$ as given 
	in~\req{PB_k_term}. For notational consistency, we now denote the initial time with $\tb$, 
	i.e. the input is applied over $[\tb,T]$ 
	\begin{align}
		\big( \Volt_A^k &~\Volt w \big)(t) 
		=
		\smint{\tb}{t} \big( \Volt_A^k\big) (t,\tau_1) \lb \smint{\tb}{\tau_1} w(\tau) ~d\tau \rb d\tau_1   	\nonumber	\\
		&= 
			\smint{\tb}{t}
			\smint{0}{T}\cdots \smint{0}{T} 
			 A(\tau_k)\cdots A(\tau_2)~  \heavi_{t,\tau_k,\ldots,\tau_{1},\tb}  
			 	~d\tau_2 \cdots d\tau_{k}  	~A(\tau_1)  
				\lb  \smint{0}{T} \heavi_{\tau_1,\tau,\tb} ~w(\tau) d\tau \rb   d\tau_1 
																			\nonumber	\\
		&= 
			\smint{\tb}{t}  \smint{0}{T} \cdots \smint{0}{T} \smint{0}{T} 
			 A(\tau_k)\cdots A(\tau_{1})~  \heavi_{t,\tau_k,\ldots,\tau_{1},\tau,\tb}  
			 	~d\tau_1 \cdots d\tau_{k}   ~w(\tau) ~d\tau , 							\nonumber	\\
		&= 
			\smint{\tb}{t} \lb \smint{\tau}{t} \cdots \smint{\tau}{\tau_2} 
			 A(\tau_k)\cdots A(\tau_{1})
			 	~d\tau_1 \cdots d\tau_{k} \rb  w(\tau) ~d\tau 							\nonumber	\\ 				
		&= 
			\smint{\tb}{t} \Phi_k(t,\tau) ~  w(\tau) ~d\tau , 				\label{Phi_VOC_k.eq}						
	\end{align} 
	The last equality follows 
	 from the expression~\req{Phi_k_expression}  for the $k$'th term of the state transition matrix. 
	
	The total response is then given by the sum over all $k$ of the terms~\req{Phi_VOC_k} 
	\begin{align*} 
		x(t) ~&=~ \sum_{k=0}^\infty \int_\tb^t \Phi_k(t,\tau) ~w(\tau) ~d\tau 
			~=~ \int_\tb^t  \lb \sum_{k=0}^\infty \Phi_k(t,\tau) \rb ~w(\tau) ~d\tau			\\
			&=~ \int_\tb^t \Phi(t,\tau) ~w(\tau) ~d\tau, 
	\end{align*} 
	where the last equation follows from the series expression~\req{PB_series}  for the state transition matrix. 
	This is precisely the input-to-state term in the variations of constants formula~\req{VOC_LTV}.

\section{Nonlinear Equations: The Picard Iteration} 									\label{Picard.sec}

	We now consider more general systems\footnote{With very minor modifications, everything in this section applies
		equally to the more general time varying case $\xd(t)=F\big(x(t),t \big)$.}  
	of the form 
	\be
		\xd(t) ~=~ A\big( x(t) \big), 
		\hstm \hstm 
		x(0) ~=~ \xb\in\R^n, 	
	  \label{NLTV_diff.eq}
	\ee
	where we set the initial time to $t=0$  for notational simplicity. 
	In such a general setting, we will not be able to say much about solutions other than existence and
	uniqueness for certain classes of vector fields $A$. This existence and uniqueness result is sometimes 
	referred to as the Picard-Lindel\"{o}f Theorem, the heart of which is the so-called {\em Picard iteration}, 
	which is the nonlinear version of the Neumann series discussed earlier. The convergence of this 
	iteration can be shown using the (Banach) fixed point theorem. The key to this argument is a 
	$C[0,T]$ norm 
	bound between successive iterates, which is accomplished by similar arguments used for bounds 
	on the action of the Volterra operator. 
	
	The differential equation~\req{NLTV_diff}
	 can be equivalently viewed as an integral equation by integrating both sides to get 
	\be
		x(t) ~-~ x(\tb) 
		~=~ \int_\tb^t A\big( x(\tau)  \big) ~d\tau 
		\hstm \Leftrightarrow \hstm 
		x ~=~ \Volt \cA \big( x \big) ~+~ \heavi \xb , 
	 \label{NLTV_int_abst.eq}
	\ee
	where the ``Heaviside operator'' maps vectors to functions $\big( \heavi \xb \big)(t) :=\xb, ~t\in[\tb,T]$, 
	$\Volt$ is the familiar integration (Volterra) operator~\req{Volterra_def}, 
	and  $\cA$ is the nonlinear point-wise  operator 
	\[
		\big( \cA(x) \big)(t) ~:=~ A\lbb x(t) \rbb .
	\]
	Unlike the linear case, we cannot write a Neumann series of the form~\req{Neu_series_sol}  
	\[
		\lbb I - \Volt \cA \rbb^{-1} 
		\neq 
		\lbb I ~+~ \Volt \cA ~+~ (\Volt \cA)^2 ~+~ \cdots \rbb , 
	\]
	because the operator $\cA$ does not distribute over additions. On the other hand, we can still 
	make sense of the iteration~\req{picard_it}
	\be
           	\begin{aligned} 
            		x_0 ~&=~ \heavi \xb  , 		\\ 
            		x_{k+1} ~&=~ \heavi \xb  ~+~ \Volt \cA(x_k) .
            	\end{aligned} 	 
	 \label{picard_it_t.eq} 
	\ee
	This is the {\em Picard iteration} in the general nonlinear case. 
	If this iteration converges, then $\lim_{k\rightarrow\infty} x_{k+1} = \lim_{k\rightarrow\infty} x_{k} =:x$, 
	and this limit $x$ satisfies the original equation~\req{NLTV_int_abst}.
	
	The convergence of the Picard iteration is dependent on properties of the function $A$. It converges
	for some, but not others. We will first give conditions and a proof of convergence over some interval 
	$[0,\epsilon)$ near the initial condition. This will follow from a classic argument using the so-called
	Banach fixed point theorem. We will then use a refinement of this technique to show global convergence 
	over all time intervals provided that the nonlinear function $A$ has a {\em linear bound} (the so-called 
	Lipschitz bound). We then close
	with some examples demonstrating the lack of uniqueness or global existence when those 
	conditions do not hold.

	\subsection{Local Convergence and Existence} 
	
	A common method to show convergence of iterations is the {\em contraction mapping theorem}
	(also called the {\em Banach fixed point theorem}~\cite{kreyszig1991introductory}), 
	whose proof is in Appendix~\ref{CMT_proof.sec}. 
	\begin{theorem} 												\label{CMT_one.thm}
		Let $M:\sfX\rightarrow\sfX$ be a mapping on a complete metric space $\sfX$ equipped with 
		the metric $\dm{.}{.}$. If the mapping is a 
		{\em strict contraction}, i.e. if 
		\[
			\dm{M(x)}{M(y) \rome} ~\leq~ \alpha ~\dm{x}{y} , 
			\hstm\hstm 
			0\leq\alpha< 1, 
		\]
		then given any initial point $x_0\in\sfX$, the sequence of iterates $x_{k+1}:= M(x_x)$ converges 
		to a unique limit $x\in\sfX$. 
	\end{theorem} 
	
	For later comparison purposes, 
	it is instructive to examine briefly the key idea behind this theorem, which is to bound the distance 
	between successive iterates by the distance between the first two 
	iterates\footnote{The notation $M^k$ stands for the mapping $M$ composed with itself $k$ times, 
		i.e. $M^2:=M\circ M$, and $M^k=M\circ \cdots \circ M$, $k$ times. Thus $x_k= M^k(x_0)$.} 
	\begin{align} 
		\dm{x_k}{x_{k+1}} 
		~&=~ \dm{M^k(x_0) \rom}{M^k(x_1)} 						\nonumber		\\
		~&\leq~ \alpha ~\dm{M^{k\sm1}(x_0) \rom}{M^{k\sm1}(x_1)} 
		~\leq~ \cdots 
			~\leq~ \alpha^k ~\dm{x_0} {x_1}.
	\label{suc_ite_bound.eq}
	\end{align}
	Since $\alpha <1$, the bound on the distance between successive iterates  
	$\lcb \alpha^k \rcb$ is a decaying geometric sequence. This can be shown to imply that the sequence 
	of iterates $\lcb x_k\rcb$ is Cauchy, and therefore convergent in the complete metric space $\sfX$. The details are
	in Appendix~\ref{CMT_proof.sec}.

	To use this theorem to examine convergence of the Picard iteration~\req{picard_it_t}, we need a suitable 
	metric on functions for which the iteration is a strict contraction. A convenient choice of metric 
	(though not the only possible choice) is in the Banach space $C[0,T]$ of 
	continuous functions equipped with the maximum norm 
	\be
		C[0,T] ~:=~ \lcb f:[0,T]\rightarrow\R^n; ~ f~\mbox{continuous}, \|f\|_\infty := \max_{t\in[0,T]} \|f(t)\|_\rmv \rcb , 
	 \label{COT_def.eq}
	\ee
	where $\|.\|_\rmv$ is any vector 
	norm\footnote{Note the since all norms on $\R^n$ are equivalent, the {\em set} $C[0,T]$ is independent 
	 	of choice of norm $\|.\|_v$. However, the norm $\|.\|_\infty$ of a function does depend on the 
		choice of the vector 
		norm.  We suppress this dependence in our notation. 
		} 
	on $\R^n$. In the iteration~\req{picard_it_t}, the mapping $M$ is 
	\be
		M(x)~:=~ \heavi \xb ~+~ \Volt\cA(x). 
	  \label{M_mapping.eq}
	\ee
	Since $C[0,T]$ is a vector space, the metric is given by the norm of the difference, and therefore 
	\begin{align} 
		&	\hspace{-2em} 
		\dm{M(x)}{M(y) \rome} 
		~=~ \left\| M(x) - M(y)  \right\|_\infty 
		~=~ 
		\left\| \heavi \xb + \Volt\cA(x) ~-~ \lb \heavi \xb + \Volt\cA(y) \rb \right\|_\infty 			\nonumber	\\
		&=~ 
		\left\|  \Volt \lbb \cA(x) - \cA(y) \rbb \right\|_\infty 
		~=~ 
		\sup_{t\in[0,T]} 	\left\| \smint{0}{t} A \lbb x(\tau) \rbb - A\lbb y(\tau)  \rbb  d\tau  \right\|_\rmv	\nonumber	\\
		 &\leq
		\sup_{t\in[0,T]} 	 \smint{0}{t} \left\| A \lbb x(\tau) \rbb - A\lbb y(\tau)  \rbb   \right\|_\rmv d\tau 
		= 
			 \smint{0}{T} \left\| A \lbb x(\tau) \rbb - A\lbb y(\tau)  \rbb   \right\|_\rmv d\tau , 
																\label{F_sup_bound.eq}
	\end{align} 
	where the last equality follows from the integrand being a non-negative function, and therefore the supremum 
	is achieved at $t=T$. 
	Now if $M$ were to be a contraction mapping, we need  to somehow bound the vector norm in the last expression
	by the vector norm $\|x(t)-y(t)\|_\rmv$. If $A$ were a linear mapping, its induced norm would give that bound. 
	If a general $F$ had such a {\em linear bound}, we could still make the same argument. This leads to the 
	following definition. 
	\begin{definition} 
		A function $A:\R^n\rightarrow\R^n$ is called {\em globally 
		Lipschitz} (or simply {\em Lipschitz})  with Lipschitz constant $\lba>0$  if 
		\be
			\forall ~x,y\in\R^n, 
			\hstm \hstm
			\| A(x) - A(y) \|_\rmv 
			~\leq~ 
			\lba ~ \|x-y\|_\rmv . 
		  \label{Lip_bound.eq}
		\ee
	\end{definition} 
	
	Note that since all vector norms on $\R^n$ are equivalent, the definition above is independent of the 
	choice of vector norm $\|.\|_\rmv$. Perhaps a more descriptive name for this property is to say that 
	the non-linear function $A$ is 
	{\em incrementally linearly bounded}\footnote{A standard linear bound would be of the form $\|F(x)\|\leq\lba \|x\|$. 
		The bound~\req{Lip_bound} is a linear bound, but on increments (differences).}. 
	Note that any linear mapping $x\mapsto Ax$ on $\R^n$ is Lipschitz, with 
	Lipschitz constant being the matrix norm of $A$ induced by the chosen vector norm on $x$. 
	
	Now returning to the bound~\req{F_sup_bound} and assuming that $A$ is Lipschitz 
	\begin{align*} 
		\hspace{-1em}
		 \smint{0}{T} \left\| A \lbb x(\tau) \rbb - A \lbb y(\tau)  \rbb   \right\|_\rmv d\tau 
		~&\leq~		
			\lba~ \smint{0}{T} \left\| x(\tau) -  y(\tau)    \right\|_\rmv d\tau 				\\
		~&\leq~		
			\lba~T~\sup_{\tau\in[0,T]}  \left\| x(\tau) -  y(\tau)    \right\|_\rmv d\tau 
			~=~ \lba~T ~\|x-y\|_\infty .
	\end{align*} 
	Combining this last bound with~\req{F_sup_bound}, we conclude that the mapping 
	$M(.):=\heavi\xb+\Volt\cA(.)$ in~\req{M_mapping}
	has the bound 
	\be
		\| M(x) - M(y) \|_\infty  ~\leq~ \lba ~T~ \|x-y\|_\infty . 
	  \label{MxMy_inf_bound.eq}
	\ee
	for any $x,y\in C[0,T]$. Clearly if we choose $T<1/\lba$, then $M$ is strictly contractive, and the contraction 
	mapping theorem implies that the Picard iteration converges on any interval $[0,\epsilon]$ provided 
	$\epsilon < \min\lcb {1}/{\lba},T\rcb$.

	\subsection{Global Convergence and Existence}

	The previous argument implied that unique solutions can  only be guaranteed to  exists on 
	proper subintervals of  
	 $[0,\min\lcb {1}/{\lba},T\rcb]$. This seems rather unsatisfactory as the interval can
	become arbitrarily small as $\lba$ becomes large. In fact, the argument we just presented is 
	unnecessarily conservative. To appreciate this, consider the linear case, where $M$ becomes 
	the mapping $\Volt_A$ from Section~\ref{Form_STM.sec}, 
	and the Picard iteration is just the Neumann series. 
	Demanding that $\Volt_A$ be a contraction mapping is equivalent to demanding that its 
	induced norm $\left\| \Volt_A \right\|<1$. However, because of the {\em causality}  
	property of the Volterra operator, we were earlier able to show that $\Volt_A^k$ converges to zero 
	in a way that insures the absolute summability of the Neumann series  even if 
	 $\left\| \Volt_A \right\|<1$ did not hold. 	
	They key condition was 
	not $\left\| \Volt_A \right\|<1$, but rather that $\left\| \Volt_A^k \right\|$ be a summable (in $k$) sequence. 
	This leads us to state a better version of the fixed point theorem. 
	\begin{theorem} 										\label{CMT_two.thm}
		Let $M:\sfX\rightarrow\sfX$ be a mapping on a complete metric space $\sfX$ equipped with 
		the metric $\dm{.}{.}$. If  the iterated distances are {\em summable for each $x$, $y$}, i.e. 
		\be
			\sum_{k=0}^\infty \dm{M^k(x)}{M^k(y) \romn} ~<~\infty,
			\hstm\hstm 
			x,y\in\sfX 
		  \label{summable_iterates_thm.eq}
		\ee
		then given any initial point $x_0\in\sfX$, the sequence of iterates $x_{k+1}:= M(x_x)$ converges 
		to a unique limit $x\in\sfX$. 
	\end{theorem} 
	It is useful to contrast this theorem with the contraction mapping theorem~\ref{CMT_one.thm}. 
	Recall that the basic idea of the latter is the bound~\req{suc_ite_bound}, which demands 
	that $\lcb \dm{x_k}{x_{k+1}} \rcb$ be a geometric sequence. The present theorem however is equivalent 
	to only requiring  that
	the sequence  $\lcb \dm{x_k}{x_{k+1}} \rcb$ be summable (see Appendix~\ref{CMT_proof.sec}). 
	Clearly a geometric sequence is summable, but that 
	is too stringent of a requirement if all one needs is summability. A sequence that can increase initially, but 
	eventually decrease at a rate that makes it summable would be admissible for Theorem~\ref{CMT_two.thm}, 
	but not for Theorem~\ref{CMT_one.thm}. 
	
	Now our goal is to show that the summability condition~\req{summable_iterates_thm} holds for the mapping 
	$M(x)=\heavi \xb + \Volt\cA(x)$. 
	To this end, we revisit the bounds~\req{F_sup_bound} and show how they can be significantly tightened. 
	The key is the ``asymptotic nilpotence'' property of the Volterra operator, which the reader should note was 
	not used in the local existence arguments. 
	
	We adopt the following notation that will make  subsequent arguments simpler to state. 
	\begin{itemize}
		\item 
			Given a vector-valued function $g:[0,T]\rightarrow\R^n$, we define its time-varying norm function
			(denoted $|g|$)  by 
			\[
				|g|(t) ~:=~ \|g(t)\|,  
			\]
			for any vector norm $\|.\|$. Of course $|g|$ depends on the specific vector norm chosen, but we suppress
			this from the notation since it will be irrelevant to the arguments we need here. 
		\item 
			Let $g,f:[0,T]\rightarrow\R$ be any two scalar-valued functions. We write $g\leq f$ if they satisfy this 
			inequality pointwise in $t$
			\[
				g~\leq~ f 
				\hstm \hstm \Leftrightarrow \hstm\hstm 
				g(t)~\leq~ f(t) , ~~ t\in[0,T] . 
			\]
		\item 
			For two vector-valued functions $f,g:[0,T]\rightarrow\R$, a combination of the above two definitions allows 
			for writing bounds of the form 
			\be
				|g|~\leq~|f| 
				\hstm\Leftrightarrow\hstm 
				\|g(t) \| ~\leq~ \|f(t)\| , 
				\hstm\hstm 
				t\in[0,T]. 
			  \label{gf_bound_point.eq}
			\ee
			Note that this is saying much more than $\|g\|_\infty\leq\|f\|_\infty$. The latter means 
			\be
				\sup_{t\in[0,T]} \|g(t)\| ~\leq~ \sup_{t\in[0,T]} \|f(t)\| .
			  \label{gf_bound_max.eq}
			\ee
			Clearly~\req{gf_bound_point} implies~\req{gf_bound_max}, but the former encodes a  more detailed 
			 comparison of the two functions $g$ and $f$. For  contraction mapping we only 
			 used bounds like~\req{gf_bound_max}.
			 We will need pointwise (in $t$) bounds like~\req{gf_bound_point} for the summability criterion. 
		\item 
			With this notation, we can write the following inequality involving the integration operator 
			\be
				\left| \Volt g \right| ~\leq~ \Volt |g| 
				\hstm \hstm \Leftrightarrow \hstm\hstm 
				\left \| \smint{0}{t} g(\tau) ~d\tau \right\| ~\leq~ \smint{0}{t} \| g(\tau) \| ~d\tau , 
				\hstm 
				t\in[0,T]. 
			  \label{Volt_bound_t.eq}
			\ee
		\item 
			If $A$ is Lipschitz with constant $\lba$, then we can write the following bounds in a compact notation 
			\[
				|A(x)-A(y)| ~\leq~ \lba ~|x-y| 
				\hstm \Leftrightarrow \hstm 
				\left\| A\lbb x(t) \rbb - A \lbb y(t) \rbb \right\| ~\leq ~\lba ~ \left\| x(t) - y(t) \right\| , 
				\hsom t\in[0,T] . 
			\]
	\end{itemize} 
	
	We now return to improving the bounds, and begin with the single-step bound 
	\begin{align*} 
		\lbar M(x) - M(y) 		\romn		\rbar
		~&=~ 
			\lbar \heavi \xb + \Volt\cA(x) - \lb \heavi \xb + \Volt\cA(y) \rb 	\romn	\rbar				
			~=~ 
			\lbar \Volt \lbb \cA(x) - A(y) \rbb	\rbar								\\
		&\stackrel{1}{\leq}~  \Volt \lbar \cA(x) - \cA(y)		\romn 	\rbar			
			~\stackrel{2}{\leq}~ 	\lba~ \Volt \lbar x -y \rbar			,			
	\end{align*} 
	where $\stackrel{1}{\leq}$ follows from~\req{Volt_bound_t} and $\stackrel{2}{\leq}$ is the Lipschitz bound. 
	For the sake of clarity, we expand this in detail for the reader that has not yet digested the new notation
	\[
		\left\| \lb  M(x) - M(y) 		\rom		\rb \!(t) \right\| 
		~\leq~ 
		\lba~ \smint{0}{t} \| x(\tau) - y(\tau) \| ~d\tau . 
	\]
	Note that this is a pointwise (in $t$) bound, and is a much better bound than~\req{MxMy_inf_bound}.	
	Now a similar bound for applying $M$ twice 
	\begin{align*} 
		\lbar M^2(x) - M^2(y)\rbar
		~&=~ 
			\lbar \heavi \xb + \Volt\cA \big(  \heavi\xb + \Volt\cA(x)  \big) 
			 	- \lb \heavi \xb + \Volt\cA \big( \heavi \xb + \Volt\cA(y) \big) \rb 	\rbar	\\					
		~&=~ 
			\lbar \Volt \lb  \cA \big( \heavi \xb + \Volt\cA(x)  \big) 
			 	-  \cA \big( \heavi \xb + \Volt\cA(y) \big)  	 \rb		\rbar				\\
		~&\leq~ 
			 \Volt \lbar  \cA \big( \heavi \xb + \Volt\cA(x)  \big) 
			 	-  \cA \big( \heavi \xb + \Volt\cA(y) \big)  			\rbar				\\
		~&\leq~ 
			\lba~ \Volt \lbar    \Volt\cA(x) 	-  \Volt\cA(y)  			\rbar
			~\leq~ 
				\lba^2~ \Volt^2    \!  \lbar x -  y  	\rbar			
	\end{align*} 
	We can therefore conclude, and verify by induction, that for any power $k$
	\be 
		\lbar M^k(x) - M^k(y)\rbar
		~\leq~ 
		\lba^k~ \Volt^k    \!  \lbar x -  y  	\rbar		. 
	  \label{MV_bound.eq}
	\ee
	A tighter bound on the last quantity is obtained by exploiting 
	the asymptotic nilpotence property of $\Volt$ 
	\begin{align*} 
		\hspace*{-1em} 
		\left\| \Volt^k      \lbar x -  y  	\rbar	\right\|_\infty
		~&=~ 
		\sup_{t\in[0,T]} \int_0^t \cdots \int_0^{\tau_2} \|x(\tau_1)-y(\tau_1)\| ~d\tau_1 \cdots d\tau_k 		\\
		~&\leq~ 
		 \int_0^T \cdots \int_0^{\tau_2}  ~d\tau_1 \cdots d\tau_k 
		 ~	\lb \sup_{\tau\in[0,T]} \|x(\tau)-y(\tau)\| \rb
		 &=~ \frac{T^k}{k!} ~\|x-y\|_\infty
	\end{align*} 
	The reader should compare this with the calculation~\req{Cauchy_rep_int} which involved the same 
	iterated integrals. 
	Finally, the pointwise bound~\req{MV_bound} implies 
	\[
		\left\| M^k(x) - M^k(y) \right\|_\infty 
		~\leq~ 
		\lba^k ~\left\| \Volt^k      \lbar x -  y  	\rbar	\right\|_\infty
		~\leq~ 
		\lba^k ~\frac{T^k}{k!} ~\|x-y\|_\infty .
	\]
	This sequence is clearly summable, and the sum can in fact be expressed as 
	\be
		\sum_{k=0}^\infty \left\| M^k(x) - M^k(y) \right\|_\infty 
		~\leq~ 
		\lb  \sum_{k=0}^\infty	\lba^k ~\frac{T^k}{k!}\rb  ~\|x-y\|_\infty
		~=~ 
		e^{\lba T} ~\|x-y\|_\infty .
	  \label{NL_good_bound.eq}
	\ee
	Thus by Theorem~\ref{CMT_two.thm}, the Picard iteration converges for any Lipschitz constant $\lba<\infty$ and any 
	interval length $T<\infty$. We summarize this formally. 
	\begin{theorem} 											\label{glob_exist.thm}
		Consider the system 
		\[
			\xd(t) ~=~ A\lbb x(t) \rbb, 
			\hstm\hstm 
			x(0)=\xb\in\R^n, 
		\]
		where $A:\R^n\rightarrow\R^n$ is globally Lipschitz.  For any interval length $T$, 
		any initial condition $\xb\in\R^n$, 
		this system 
		has a unique solution over $[0,T]$. 
	\end{theorem} 
	
	We note that the bound~\req{NL_good_bound} is tight in the sense that the simple scalar linear system 
	\[
		\xd(t) ~=~ \lba ~x(t), 
	\]	
	has as solution $x(t) ~=~ e^{\lba t} ~x(0)$, which grows in time proportionally to the bound~\req{NL_good_bound}.

	\subsection{Examples}

	\begin{example} 
		Consider the scalar nonlinear system
		\be
			\xd(t) ~=~ x^2(t).
		   \label{orignon_early.eq}	
		\ee
		This scalar differential equation is  solvable by ``separation of variables'' and direct integration
		\begin{align*} 
			\hspace*{-1em} 
			\xd(\tau) = x^2(\tau)
			\hsom &\Leftrightarrow \hsom 
			\frac{1}{x^2} \frac{dx}{d\tau} = 1 
			\hsom \Rightarrow \hsom 
			\smint{x(0)}{x(t)} \frac{1}{x^2} dx = \smint{0}{t} d\tau
			\hsom \Rightarrow \hsom 
			-\left( \frac{1}{x(t)} - \frac{1}{x(0)} \right) = t			\\
			&\Rightarrow\hsom
			x(t) ~=~ \frac{x(0)}{1-x(0) ~t}
		\end{align*} 
		This equation has a solution for small initial times, but it
		has  the interesting feature of ``finite escape 
		time'' as $t$ approaches ${1/x(0)}$. That is the solution asymptotes to infinity as $t\rightarrow 1/x(0)$. 
		The larger the initial condition, the shorter is the time interval over which the solution is possible. 
		The right hand side of~\req{orignon_early} is not globally Lipschitz, and therefore 
		Theorem~\ref{glob_exist.thm} 
		does not apply. There is however a notion of {\em locally Lipschitz} systems for which only local existence 
		can be guaranteed, with the time interval of existence being dependent on the initial condition. 
		This is the situation with this example. 
	\end{example}

	\begin{example} 									\label{nonunique_square_root.example}
		In this example, solutions need not be unique. This usually happens when the vector field $F$ has infinite 
		derivatives. Consider the scalar system 
		\[
			\xd(t) ~=~ \sqrt{|x(t)|}, 
			\hstm\hstm 
			x(0) = 0 . 
		\]
		Clearly $x(t)=0$,  $t\geq0$ is a solution, but there are others. By separation of variables again
		\begin{align*} 
			\xd(\tau) = \sqrt{x(\tau)}
			\hsom &\Leftrightarrow \hsom 
			x^{\sm\frac{1}{2}} \frac{dx}{d\tau} = 1 
			\hsom \Rightarrow \hsom 
			\left. 2x^{\frac{1}{2}} \right|_{x(0)}^{x(t)} = t
			\hsom \Rightarrow \hsom 
			x(t) = \frac{t^2}{4}		.
		\end{align*} 
		Thus we have found two different solutions from the same initial conditions. Note that these solutions are 
		valid for all $t\in[0,\infty)$. Therefore, the non-uniqueness phenomenon is a separate one from the finite-escape-time 
		phenomenon. 
	\end{example}

	\subsection{Modeling Implications of Existence and Uniqueness} 				\label{commentary.sec}
		
		Both of the examples above highlight an important issue in mathematical modeling of physical systems. 
		We generally believe that given enough information about a physical system, we can construct
		a mathematical model (e.g. a differential equation) that predicts the future behavior of the system given 
		a fully accurate (infinite precision) description of initial 
		conditions\footnote{The discussion here is unrelated to the phenomenon of ``chaos'', which involves 
			sensitive dependence on initial conditions. There are many chaotic systems with solutions that are 
			guaranteed to exists from all initial conditions and are unique.}. 
		All models however are approximate, and no one believes that their mathematical model of any physical 
		phenomena is fully accurate in all regimes\footnote{Those who do not realize that, are usually writing science 
		fiction,  contemplating Schr\"{o}dinger's cat, or some other similar speculation.}. 
		If we have finite-escape-time, this means that quantities (e.g. velocities, pressures, etc.) are becoming so large
		that the mathematical model is no longer fully valid. If we have differential equations that are not locally 
		Lipschitz (such as Example~\ref{nonunique_square_root.example} above), 
		this means that derivatives (usually forces in mechanical models) 
		become arbitrarily sensitive to small changes in the state. This is again a regime where the mathematical 
		model breaks down, and no longer accurately represents the physical world.  
	
		The theme of the above remarks is that non-uniquness or lack of existence of solutions is  not a {\em mathematical }
		difficulty, but rather a {\em mathematical modeling} difficulty. One can come up with equations and mathematical 
		constructs that do all kinds of fantastical things. The question is whether these are good mathematical models
		of the physical world. It seems like a natural minimal requirement that a mathematical model should posses
		the property of existence and uniqueness of solutions.

\bibliographystyle{IEEEtran}
\bibliography{22_SS_solutions}

\begin{thebibliography}{1}
\providecommand{\url}[1]{#1}
\csname url@samestyle\endcsname
\providecommand{\newblock}{\relax}
\providecommand{\bibinfo}[2]{#2}
\providecommand{\BIBentrySTDinterwordspacing}{\spaceskip=0pt\relax}
\providecommand{\BIBentryALTinterwordstretchfactor}{4}
\providecommand{\BIBentryALTinterwordspacing}{\spaceskip=\fontdimen2\font plus
\BIBentryALTinterwordstretchfactor\fontdimen3\font minus
  \fontdimen4\font\relax}
\providecommand{\BIBforeignlanguage}[2]{{%
\expandafter\ifx\csname l@#1\endcsname\relax
\typeout{** WARNING: IEEEtran.bst: No hyphenation pattern has been}%
\typeout{** loaded for the language `#1'. Using the pattern for}%
\typeout{** the default language instead.}%
\else
\language=\csname l@#1\endcsname
\fi
#2}}
\providecommand{\BIBdecl}{\relax}
\BIBdecl

\bibitem{kailath1980linear}
T.~Kailath, \emph{Linear systems}.\hskip 1em plus 0.5em minus 0.4em\relax
  Prentice-Hall Englewood Cliffs, NJ, 1980, vol. 156.

\bibitem{antsaklis1997linear}
P.~J. Antsaklis and A.~N. Michel, \emph{Linear systems}.\hskip 1em plus 0.5em
  minus 0.4em\relax Springer, 1997, vol.~8.

\bibitem{hespanha2018linear}
J.~P. Hespanha, ``Linear systems theory,'' in \emph{Linear Systems
  Theory}.\hskip 1em plus 0.5em minus 0.4em\relax Princeton university press,
  2018.

\bibitem{riesz2012functional}
F.~Riesz and B.~S. Nagy, \emph{Functional analysis}.\hskip 1em plus 0.5em minus
  0.4em\relax Courier Corporation, 2012.

\bibitem{kreyszig1991introductory}
E.~Kreyszig, \emph{Introductory functional analysis with applications}.\hskip
  1em plus 0.5em minus 0.4em\relax John Wiley \& Sons, 1991, vol.~17.

\bibitem{bernard2013interpolation}
C.~Bernard, ``Interpolation theorems and applications,'' \emph{University of
  Chicago REU}, 2013.

\end{thebibliography}

		\newpage 
		
\appendix

\begin{center} 
	{\bf \LARGE Appendix} 
\end{center}

	\section{Convergence of the Neumann and the Peano-Baker Series} 		\label{Neu_conv.app}
	
	 For clarity, we first show the convergence 
	of the Neumann series for the Volterra operator alone 
	\[
		\big( I - \Volt \big)^{-1} ~=~ \sum_{k=0}^\infty \Volt^k , 
	\]
	and then show it for the Peano-Baker series. 
	Let $\|\Volt\|_\rmi$ denote any induced operator norm (e.g. the $\LP{1}[0,T]$ or $\LP{\infty}[0,T]$ induced norms). 	
	One commonly used sufficient condition for the convergence of this 
	series in the operator norm is $\|\Volt\|_\rmi<1$, which then renders $\|\Volt\|_\rmi^k$ a geometrically 
	convergent series. However, this condition is far from necessary, and a better condition is the 
	summability of the series $\left\| \Volt^k \right\|_\rmi$. This is the case here because of the ``lower-triangular'' structure of $\Volt$
	even though in general we could have $\|\Volt\|_\rmi>1$. 
	
	Recall the expression~\req{Volterra_k}  	for the kernel function of the $k$'th power of $\Volt$
	\be
		\Volt^k(t,\tau) ~=~ \frac{(t-\tau)^{k-1}}{(k-1)!} ~ \heavi(t\sm\tau) . 
	  \label{Vk_kernel.eq}
	\ee
	The  $\LP{1}[0,T]$ or $\LP{\infty}[0,T]$ induced norms of an operator are easy to bound from its kernel 
	representation, so we denote either of  those particular induced norms by $\|.\|_\rmi$ here. 
	This norm is bounded by the maximum of 
	the kernel function over $[0,T]\times[0,T]$. For the operator $\Volt^k$ 
	\be
		\left\| \Volt^k \right\|_\rmi ~\leq~ \sup_{t,\tau\in[0,T]} \left| \frac{(t-\tau)^{k-1}}{(k-1)!} ~ \heavi(t\sm\tau) \right|
		~=~ \frac{T^{k-1}}{(k-1)!}~ . 
	  \label{Vk_ind_bound.eq}
	\ee
	Thus the series converges absolutely in the induced operator norm (on both $\LP{1}[0,T]$ and $\LP{\infty}[0,T]$). 
	The Riesz-Thorin convexity theorem~\cite{bernard2013interpolation}
	 then implies convergence in the induced operator norm on $\LP{p}[0,T]$ 
	for any $p\in[1,\infty]$. 
	
	For the general time-varying case, the only assumption needed is that the function $A(.)$ is bounded on 
	bounded intervals. 
	Recall that the Peano-Baker series~\req{Neumann_LTV_hI}  for the state transition matrix is
	\be
		\Phi ~=~ \big( I  - \Volt _A \big)^{-1}~   \heavi I 
		~=~\lb \sum_{k=0}^\infty \Volt_A^k  \rb ~\heavi I. 
	  \label{Phi_hI_series.eq}
	  \ee
	  We will show that $\sum_{k=0}^\infty \Volt_A^k$ is convergent in the operator norm on $C[0,T]$. 
	  Since $\heavi I$ is a constant function on $[0,T]$, then the expression above converges to a function in 
	  $C[0,T]$. As in the previous case of the Volterra operator, the $C[0,T]$-induced norm (denoted by
	  $\left\| \Volt_A^k\right\|_\rmi$) is bounded from above by the maximum absolute value of the kernel 
	  function over $[0,T]\times[0,T]$. Now compute 
	  \begin{align} 
		\left\| \Volt^k_A \right\|_\rmi 
			~&\leq~ 
				\sup_{t,\tau_1\in[0,T]} \left\|
					\smint{0}{T}\cdots \smint{0}{T} 
					 A(\tau_k)\cdots A(\tau_{2})~  \heavi_{t,\tau_k,\ldots,\tau_{2},\tau_1,\tau}  
			 		~d\tau_2 \cdots d\tau_{k}  	~~A(\tau_1)  \right\|						\nonumber	\\
			&\leq~ 
					\lb \sup_{t\in[0,T]} \left\| A(t) \right\|  	\rb
					~\left|  \smint{\tau}{t}\cdots \smint{\tau}{\tau_3} 
					d\tau_2 \cdots d\tau_{k}  \right| 								\label{VAk_bound_o.eq}	\\
			&=~ 
					\lb \sup_{t\in[0,T]} \left\| A(t) \right\|  	\rb
					~\left| \Volt^k(t,\tau) \right| 
			~\leq~ 	\lb \sup_{t\in[0,T]} \left\| A(t) \right\|  	\rb
					\frac{T^{k-1}}{(k-1)!}	.									\label{VAk_bound_t.eq}
	  \end{align}
	 The equality in the last line comes from observing that the integral in~\req{VAk_bound_o} is precisely 
	  the kernel function~\req{Vk_kernel}  of the operator $\Volt^k$. This kernel in turn has the 
	  bound~\req{Vk_ind_bound}. This last bound guarantees that the series converges absolutely. 
	  
	  We finally 
	  note that a similar argument to the above can be used to show convergence of the series in the induced
	  operator norm over $C^m[0,T]$, the (Sobolev) Banach space of $m$ times differentiable functions, with the maximum 
	  norm of the first $m$ derivatives. This then shows that for the series~\req{Phi_hI_series}, all first $m$ derivatives 
	  converge uniformly, and therefore converge to an $m$-times continuously differentiable function. Since $m$ 
	  can be any integer, then the state transition matrix (for a finite-dimensional system) is smooth (infinitely 
	  differentiable).

	\section{The Peano-Baker Series in the Commutative Case} 							\label{PBS_commutative.appen}
	
	The proof of Lemma~\ref{PBS_commutative.lemma} relies on the following observation
	\begin{align*}
		~
		\smint{0}{t} A(\tau) \smint{0}{\tau} A(r) ~dr ~d\tau ,
            		~=~ 
			\smint{0}{t} \dot{F}(\tau) ~F(\tau) ~d\tau, 
            		\hstm \hstm 
            		\mbox{where}~~ 
            		F(\tau) &:= \smint{0}{\tau} A(r) ~dr 				\\ 
			\mbox{and therefore}~~ 
			\dot{F}(\tau) &= A(\tau) , ~F(0)=0. 	
	\end{align*} 
	This indicates that we could use the integration-by-parts formula for matrix-valued functions to 
	simplify this expression. This simplification is possible  provided that 
	$F(\tau)$ and $\dot{F}(\tau)$ commute. 
	First, the  commutativity of the family $\lcb A(t), ~t\in[0,T] \rcb$ implies that $A(t)$ commutes with its 
	integrals 
	\begin{align*} 
		A(t) \lb   \smint{0}{t} A(\tau) ~d\tau \rb
		&= 
		 \lb   \smint{0}{t} A(t)~A(\tau) ~d\tau \rb
		= 
		 \lb   \smint{0}{t}A(\tau) ~ A(t)~d\tau \rb
		= 
		\lb  \smint{0}{t} A(\tau) ~d\tau \rb  A(t) 		\\
		 \Rightarrow \hstm 
		\dot{F}(t) ~F(t) ~&=~ F(t) ~ \dot{F}(t). 
	\end{align*} 
	Now observe that we can use integration-by-parts on matrix-valued functions 
	\begin{align*}
		~\smint{0}{t} \dot{F}(\tau) ~F(\tau) ~d\tau,       						
		~&=~  
			\left. F(\tau) F(\tau) \rom \right|_0^t - \smint{0}{t}  F(\tau) ~\dot{F}(\tau) ~d\tau 
			~\stackrel{\txtcm{1}}{=}~
			 F^2(t) - \smint{0}{t}  \dot{F}(\tau) ~{F}(\tau) ~d\tau					\\
		\hspace*{-1.5em}
		\Rightarrow~~ 
			\smint{0}{t} \dot{F}(\tau) ~F(\tau) ~d\tau 
			&=~ 
			\frac{1}{2} ~F^2(t) , 								
	\end{align*} 
	where we used the commutativity of $\dot{F}$ and $F$ in $\stackrel{\txtcm{1}}{=}$. 
	
	By induction, we can show that 
	\begin{align*} 
		\Phi_k(t) ~&:=~ 
		\smint{0}{t} A(\tau_k) \smint{0}{\tau_k} A(\tau_{k\sm1})  \cdots  \smint{0}{\tau_2} A(\tau_1) ~d\tau_1  \cdots d\tau_k 
			~=~ \smint{0}{t} A(\tau_k) \Phi_{k-1}(\tau_k) ~d\tau_k 								\\
		&=~ 
			\frac{1}{k!} \lb \smint{0}{t} A(\tau) ~d\tau \rb^k  
			~=~ \frac{1}{k!} ~F^k(t).
	\end{align*} 
	Indeed, assume the statement is true for $k-1$, then the above expression for $\Phi_k$ 
	is\footnote{After the first line of this calculation, the dependence on $\tau$ is suppressed for notational simplicity.} 
	\begin{align*} 
		\Phi_k(t) = 
		\smint{0}{t} A(\tau)~ \Phi_{k\sm1}(\tau) ~\tau 
			~&=~ \frac{1}{(k\sm1)!} ~ \smint{0}{t} \dot{F}(\tau) ~ F^{k\sm1}(\tau)  ~d\tau 					\\
		&=~ 
			\frac{1}{(k\sm1)!} \lb  \left.  F  F^{k\sm1} \right|_0^t - (k\sm1) 
			 \smint{0}{t} F \lb \dot{F} F^{k\sm2} +F^{k\sm2} \dot{F} \rb   d\tau	\rb					\\
		\Rightarrow ~~ 
			 \smint{0}{t} \dot{F} F^{k\sm1} ~   d\tau
		~&=~ 
			F^{k} - (k\sm1) 
			 \smint{0}{t} \dot{F} F^{k\sm1} ~   d\tau	
			 																			\\
		\Rightarrow ~~ 
				 \smint{0}{t} \dot{F} F^{k\sm1} ~   d\tau
		~&=~ 
			\frac{1}{k}      F^{k} 	
			\hstm \Rightarrow \hstm 
			\frac{1}{(k\sm1)!} \smint{0}{t} \dot{F} F^{k\sm1} d\tau = \frac{1}{k!} F^k,
	\end{align*} 
	where we used the fact that if $\dot{F}$ commutes with $F$, then it commutes with all powers of $F$. 
	
	Finally, the state transition matrix is given from the series by 
	\[
		\Phi(t,0) ~=~ \sum_{k=1}^\infty \Phi_k(t) ~=~  \sum_{k=1}^\infty \frac{1}{k!}\lb \smint{0}{t} A(\tau)~d\tau \rb^k
			~=~ \exp\lb \smint{0}{t} A(\tau)~d\tau \rb .
	\]

	\section{Proof of the Contraction Mapping Theorems} 					\label{CMT_proof.sec}

	First we consider Theorem~\ref{CMT_one.thm}. As explained in~\req{suc_ite_bound}, the key is that 
	the iterates satisfy 
	\[
		\dm{x_k}{x_{k+1}} ~\leq~ \alpha^k ~\dm{x_0}{x_1}. 
	\]
%
	Now let $N$ be some integer and consider the distance between $x_N$ and any element $x_n$ in the
	 ``tail'' of the sequence (i.e. $n>N$) 
	\begin{align*} 
		\dm{x_N}{x_n} 
		~&\leq~ \dm{x_N}{x_{N+1}} ~+~ \cdots ~+~ \dm{x_{n\sm1}}{x_n} 
			~\leq~ \big( \alpha^N + \cdots + \alpha^{n-1}\big) ~\dm{x_0}{x_1} 				\\
		&=~ \lb  \sum_{l=N}^{n-1} \alpha^l \rb \dm{x_0}{x_1} 
			~\leq~ \lb  \sum_{l=N}^{\infty} \alpha^l \rb \dm{x_0}{x_1} 
			~=~ \frac{\alpha^N}{1-\alpha} ~ \dm{x_0}{x_1}, 
	\end{align*} 
	where the first inequality follows from the triangle inequality. 
	Since $\alpha<1$, the last bound (which is on the entire tail) can be made as small as desired. 
	This proves that $\lcb x_k \rcb$ is Cauchy. Since $\sfX$ is a complete metric space, the sequence therefore  has a 
	unique limit. 
	
	Now for Theorem~\ref{CMT_two.thm}. The summability condition 
	\[	\textstyle 
		\sum_{k=0}^\infty \dm{M^k(x)}{M^k(y) \romn} ~<~\infty	, 
		\hstm\hstm 
		x,y\in\sfX, 
	\]
	implies the summability of the  distances between successive iterates 
	\[
		\sum_{k=0}^\infty \dm{x_k}{x_{k+1}} ~=~ \sum_{k=0}^\infty \dm{ M^k(x_0) }{M^k(x_1)	\rom } 
		~<~ \infty. 
	\]
	The fact that the series sum of  successive distances is finite implies that the sequence $\lcb x_k \rcb$ 
	is Cauchy. Indeed, summability implies that given any $\epsilon$, $\exists N$ such that 
	\[
		\sum_{k=N}^\infty \dm{x_k}{x_{k+1}} ~\leq~ \epsilon 
		\hstm \Rightarrow \hstm 
		\dm{x_{k_1}}{x_{k_2}} ~\leq~ \epsilon , ~~\mbox{for}~ k_1,k_2\geq N. 
	\]

\newpage

\section*{Exercises}

	\begin{ExInternal}												\label{Flow_PDE.ex}	

		Given the system $\xd(t) ~=~ A\blb x(t),t \brb$, show that its flow map $\Phi_{t,\tb}$ satisfies the 
		``functional'' partial differential equation 
		\[
			\frac{\partial}{\partial t} ~\Phi_{t,\tb} (x) ~=~ A \big( \Phi_{t,\tb} (x) , t \big) , 
			\hstm\hstm 
			\Phi_{\tb,\tb} (x) ~=~x . 
		\]
		
	\end{ExInternal}	
	
	

	\begin{ExInternal}														\label{STM_scalar.ex}	

		Given the scalar ($x(t)\in\R$) time-varying system 
		\[
			\xd(t) ~=~ a(t) ~x(t), 
			\hstm\hstm 
			x(0) ~=~ \xb, 
		\]		
		show that the solution is given by the formula 
		\[
			x(t) ~=~ e^{\int_0^t a(\tau) d\tau} ~\xb. 
		\]
		{\em Hint: Rewrite the equation as $\frac{d}{dt} \ln\big( x(t) \big) ~=~ \frac{\xd(t)}{x(t)} = a(t)$.} 
		
	\end{ExInternal}

	

	\begin{ExInternal}													\label{Leibniz.ex}	

		The \index{Leibniz integral rule}{Leibniz integral rule} is the fundamental theorem of calculus when the integral limits depend 
		on the differentiation variable. It states that 
		\[
			\frac{d}{dt} \int_{\llb(t)}^{\lub(t)} f(t,\tau) ~d\tau ~=~
				   \int_{\llb(t)}^{\lub(t)} \frac{\partial}{\partial t} f(t,\tau) ~d\tau
					~+~ 
					f\big( t, \lub(t) \big) ~\lub'(t) ~-~ f\big( t, \llb(t) \big) ~\llb'(t) , 
		\]
		where the lower and upper bound functions are such that $\llb(t)\leq\lub(t)$, and $\lub'$ is notation for the derivative
		of $\lub$. 
		Prove this formula by rewriting the integral as 
		\[
			 \int_{\llb(t)}^{\lub(t)} f(t,\tau) ~d\tau 
			 ~=~ 
			 \int_{-\infty}^\infty f(t,\tau) ~\heavi\big( \lub(t)-\tau\big) ~\heavi\big(\tau- \llb(t)\big) ~d\tau ,
		\]
		and using both the product rule, as well as the fact that the derivative of the 
		unit-step (Heaviside) function is the Dirac delta function. 
		
	\end{ExInternal}

	

	\end{document}